\documentclass[%
 preprint,
 onecolumn,
superscriptaddress,
footinbib, % puts footnotes as bib items
 amsmath,amssymb,
 aps,
]{revtex4-2}

\usepackage{graphicx}% Include figure files
\usepackage{dcolumn}% Align table columns on decimal point
\usepackage{bm}% bold math
\usepackage{xcolor}
\usepackage{soul}
\usepackage{setspace}
\begin{document}

%\preprint{APS/123-QED}
%\reprint{APS/123-QED}

\title{
Emergence of dissipation and hysteresis from interactions among reversible, non-dissipative units: The case of fluid-fluid interfaces
}

\author{Ran Holtzman}
\email{ran.holtzman@coventry.ac.uk}
\affiliation{Centre for Fluid and Complex Systems, Coventry University, Coventry, United Kingdom}

\author{Marco Dentz}
\affiliation{Institute of Environmental Assessment and Water Research (IDAEA), Spanish National Research Council
(CSIC), Barcelona, Spain}

\author{Marcel Moura}
\affiliation{PoreLab, The Njord Centre, Department of Physics, University of Oslo, Oslo, Norway}

\author{Mykyta Chubynsky}
\affiliation{Centre for Fluid and Complex Systems, Coventry University, Coventry, United Kingdom}

\author{Ramon Planet}
\affiliation{Departament de F\'{\i}sica de la Mat{\`e}ria Condensada, 
Universitat de Barcelona, 
%Mart\'{\i} i Franqu{\`e}s 1, 08028 
Barcelona, Spain, 
%}
% \affiliation{
and
University of Barcelona 
Institute of Complex Systems, Barcelona, Spain}

\author{Jordi Ort{\'\i}n}
\affiliation{Departament de F\'{\i}sica de la Mat{\`e}ria Condensada, 
Universitat de Barcelona, 
%Mart\'{\i} i Franqu{\`e}s 1, 08028 
Barcelona, Spain, 
%}
% \affiliation{
and
University of Barcelona 
Institute of Complex Systems, Barcelona, Spain}

%\date{\today}

\begin{abstract}
We examine the nonequilibrium nature of two-phase fluid displacements in a quasi-two-dimensional medium (a model open fracture), in the presence of localized constrictions (``defects''), from a theoretical and numerical standpoint. Our analysis predicts the capillary energy dissipated in abrupt interfacial displacements (jumps) across defects, and relates it to the corresponding hysteresis cycle, e.g. in pressure-saturation. We distinguish between ``weak'' (reversible interface displacement, exhibiting no hysteresis and dissipation) and  ``strong'' (irreversible) defects. We expose the emergence of dissipation and irreversibility caused by spatial interactions, mediated by interfacial tension, among otherwise weak defects. We exemplify this cooperative behavior for a pair of weak defects and establish a critical separation distance, analytically and numerically, verified by a proof-of-concept experiment.
\end{abstract}

%\pacs{64.70.qj, 05.40.-a, 68.35.Ct}
% 64.70.qj Dynamics and criticality
% 05.40.-a Fluctuation phenomena, random processes, noise, and Brownian motion
% 47.53.+n Fractals in fluid dynamics
% 47.56.+r Flows through porous media
% 68.35.Ct Interface structure and roughness
% 05.45.?a Nonlinear dynamics and chaos
% 05.70.Np Interface and surface thermodynamics

%\keywords{Suggested keywords}%Use showkeys class option if keyword
                              %display desired

\maketitle

\newcommand{\dbar}{d\hspace*{-0.08em}\bar{}\hspace*{0.1em}}

% \newpage
% \tableofcontents

%%%%%%%%%%%%%%%%%%%%%%%%%%%%%%%%%%%%%%%%%%%
\section{Introduction}
\label{sec:introduction}
%%%%%%%%%%%%%%%%%%%%%%%%%%%%%%%%%%%%%%%%%%%%%%%%%

Path-dependency (hysteresis) in pressure-saturation relationships during imbibition and drainage in two-phase displacements in porous media occurs in wide variety of natural and engineered processes, for instance soil moisture and geoenergy \cite{Sahimi2011,Albers2014}.
This hysteresis is largely due to individual and cooperative capillary instabilities, known as Haines jumps \cite{Haines1930}, which are inherently related to energy dissipation~\cite{Morrow1970,Berg2013}.
The ubiquity of these phenomena in various applications and the intriguing underlying physics motivated extensive experimental, numerical and theoretical studies~\cite{Berg2013,Cueto2016,Helland_WRR2021,maloy2021burst,McClure_PRE2021,PrimkulovPRL2020,Bedeaux_Kjelstrup_Entropy_2022,Holtzman2023}. 
The intrinsic complexity of porous  media and non-local pore-scale interactions makes the quantitative understanding of the precise mechanisms that lead to these phenomena a challenging task. 
Modeling approaches that account for the metastability of two-fluid configurations and for hysteresis are often based on the aggregation of individual hysteretic units (hysterons), in the so-called compartment models (e.g. \cite{Cueto2016,Helland_WRR2021}).
Recently, based on quantitative insights from systematic studies in simplified model systems that allow to isolate individual features causing hysteresis \cite{Joanny1984,Planet2020}, a novel approach that does not rely on the concept of hysterons was suggested in \cite{HoltzmanCommPhys2020}.

However, the possibility that non-hysteretic units interact cooperatively to give rise to hysteresis and dissipation remains unexplored, despite evidences from %although the interaction between defects is known to be responsible for hysteresis and dissipation in 
paradigmatic models of collective phenomena such as the Random Field Ising Model (RFIM) \cite{SethnaPRL1993,Ortin1998}.
Here we address this open question by studying the passage of a two-phase interface through localized perturbations in the capillary pressure, representing single topographic defects (gap thickness constrictions) in a 2-D medium (Hele-Shaw cell). 

%{\color{magenta} 
Two-phase fluid flow in Hele-Shaw cells---quasi-two-dimensional setups formed by two closely spaced parallel plates
\cite{deGennes1986,Soriano2002,Geromichalos2002}---shares some salient features with two-phase flows in porous and fractured media. 
%Hele-Shaw cells are quasi-two-dimensional setups formed by two closely spaced parallel plates, where fluid flow takes place in the narrow gap spacing between them. 
In both, the bulk behavior of the flow of fluids in the viscous (Stokes) regime can be described by Darcy's law~\cite{Darcy1856}, and the stability of fluid-fluid interfaces is controlled by viscosity and density contrasts between the fluids \cite{Saffman1958}. 
A richer, more realistic model system is the ``imperfect'' Hele-Shaw cell, featuring gap-thickness constrictions and expansions,
%{\color{red} protrusions} \hl{is that a proper word here? Would ``expansions'', as used elsewhere in the paper, be better?}, 
in which the interfaces are subjected to capillary instabilities akin to those observed in disordered media \cite{Joanny1984,deGennes1986}. Imperfect Hele-Shaw cells therefore stand out as an ideal playground to study two-phase displacements in disordered media.
%} 

Using this model system, we study the notion of weak (reversible i.e. non-dissipative and non-hysteretic) and strong (dissipative and hysteretic) defects. We first classify single topographic defects as weak or strong, and quantify the associated energy dissipation and hysteresis. We then demonstrate, analytically, numerically, and experimentally, the emergence of dissipation and hysteresis due to spatial interactions (surface tension) 
%due to  
%because the capillary forces of the two defects become correlated through 
%surface tension along the interface 
among individually weak defects. We show that a pair of weak defects that is non-dissipative if separated far apart becomes strong (dissipative) when brought close enough together. % as a result of  

The paper is organized as follows.
Section \ref{sec:quasistatic} provides background on quasistatic pressure-driven displacements in imperfect Hele-Shaw cells, including the mathematical statements of pressure equilibrium and energy dissipation. 
Section \ref{sec:config_single} presents the solution for the interfacial configurations 
% introduces 
% the model media, an imperfect Hele-Shaw cell~\cite{deGennes1986},
% and reviews quasistatic pressure-driven displacements 
across single defects 
% gap-thickness heterogeneities (constrictions and expansions). 
(capillary pressure distortions).
We show the functional form of the interfacial shapes, which also reveal subtle morphological differences between imbibition and drainage, and distinct between weak (reversible, non-dissipative) and strong (irreversible, dissipative) defects. 
Section \ref{sec:energy_balance} uses the above to compute an energy balance from which we establish the energy dissipated during jumps.
%to distinguish between weak (reversible, non-dissipative) and strong (irreversible, dissipative) defects. \hl{We actually first distinguish between them in} Sec.~\ref{sec:config_single}, \hl{on the basis of jumps in interface shapes, and then confirm} in Sec.~\ref{sec:energy_balance} \hl{by calculating dissipation.}
%
%We obtain the capillary-pressure vs saturation trajectories \hl{we don't ever present our results in terms of saturation, except in the videos} in quasistatic pressure-driven displacements through strong defects, and apply an energy balance 
%\hl{RP: for me it feels weird to refer to a future equation in this paragraph and I don't think we need to refer it at this point.}
%
In Section \ref{sec:cooperative_dissipation}
% is dedicated to demonstrate that energy may be dissipated also when gap-thickness heterogeneities are weak, as a result of their mutual influence through the interfacial tension. %This section begins with providing experimental evidence for this cooperative behavior. 
we consider cooperative effects in the presence of multiple defects, with a nonintuitive result---energy dissipation due to spatially-correlated interactions through interfacial tension for a pair of weak defects brought sufficiently close together. 
A theoretical analysis 
provides a critical value for the lateral separation between two weak defects that makes them collectively dissipative. 
This prediction, derived in the limit of very narrow defects, is validated against numerical simulations, and qualitatively verified experimentally.
Finally, in Sec.\ \ref{sec:discussion_conclusions} the original results of the work are briefly recalled and the main conclusions are drawn.
Specific details of the calculations, numerical simulations and experiments are provided in the Appendices.

\section{Background}
\label{sec:quasistatic}

\subsection{Model system: Imperfect Hele-Shaw cell}

Quasistatic pressure-driven displacements in imperfect Hele-Shaw cells have shown strongly nonlinear behavior and Haines jumps at pore and multipore scale, together with lack of reversibility between opposite displacement directions (imbibition and drainage) at continuum scale, hysteresis in the applied pressure vs wetting-phase saturation (PS) trajectories, and the return-point memory (RPM) property of closed partial cycles that is ubiquitous in porous media flows \cite{Planet2020,HoltzmanCommPhys2020,Lavi_PRF_2023}.
Below, we describe the theory allowing to compute the interfacial configurations and energy dissipation for the general case of an imperfect Hele-Shaw cell 
%\hl{RP: we alternate between "HS cell" and "Hele-Shaw cell", Since it is a long paper I would avoid using the contraction, but if we decide to use it, then use it across the whole manuscript}
with multiple defects (representing disordered media).
%\hl{RP: do we need to mention "disorder"? we could leave it with "multiple defects"}
In this paper we apply these concepts to displacements through (i) isolated defects (local constrictions), and (ii) a pair of defects, exposing their interaction.

\subsection{Establishing the equilibrium configurations}

The pressure balance of a two-phase interface invading an imperfect Hele-Shaw cell can be derived in quasistatic conditions \cite{HoltzmanCommPhys2020,Planet2020}. 
Without loss of generality, we assume in the following that the less wetting fluid is low-viscosity (e.g. air) and the more wetting fluid is viscous (e.g. liquid such as silicone oil).
We consider that (i) the gap thickness in the cell changes in space 
(giving rise to an extended domain of connected constrictions and expansions); (ii) the fluids are immiscible, and displacements are driven by the change of the imposed pressure $P$ at one end of the cell; and (iii) the cell is tilted in the direction that prevents the formation of viscous fingers \cite{Saffman1958}.  
With these conditions, the linearized pressure balance takes the form 
%\cite{HoltzmanCommPhys2020,Planet2020}
%
\begin{equation}
    \label{eq:press_balance}
\gamma \frac{d^2 h(x)}{d x^2} - \rho g_e h(x) + P + p_c \left[ x,h(x) \right] = 0,
\end{equation}
where $h(x)$ is the equilibrium interface position at $x$, $\gamma$ the oil-air surface tension, $\rho$ the oil density, $g_e$ is the effective gravity (which in a physical implementation could be changed by tilting the cell, $g_e =g \sin{\alpha}$, where $g$ is the gravitational acceleration and $\alpha$ the inclination angle from the horizontal). Here $p_c(x,y)$ is the perturbation in out-of-plane capillary pressure, determined by the variations in thickness ($z$). 
Here, the direction of the fluid advancement is parallel to the $y$-axis.
For simplicity, we do not account for the minute variations in hydrostatic pressure of non-wetting fluid relative to the liquid pressure of the wetting fluid. Additionally, in our quasi-static model, the pressure changes associated with the viscosity of the two fluids are also neglected. 
The first and last terms 
%$\propto d^2 h / d x^2$ and $p_c[x,h(x)]$ 
in Eq.~(\ref{eq:press_balance}) account for the linearized in-plane component of the Young-Laplace pressure jump across the interface at each site $x$ (for comparison with the exact nonlinear term see \cite{Lavi_PRF_2023}), and the out-of-plane component arising from the presence of expansions and constrictions in the cell, respectively. 
The role of the two terms is different: 
the out-of-plane component 
is responsible for the interface deformation whereas the in-plane component is a restoring force resisting the deformation. 
In the quasistatic limit (zero driving rate) displacements are driven by minute changes of $P$, separated by long time intervals required for reaching a new mechanical equilibrium $h(x)$.

%As explained in detail in Ref.~\cite{HoltzmanCommPhys2020} the pressure balance in Eq.~\eqref{eq:press_balance} 
The equilibrium configurations $h(x)$ 
could also be derived from minimizing the Hamiltonian 
% follows also from minimizing the Hamiltonian 
\begin{equation}
\label{eq:hamiltonian}
\mathcal H = \int_{-\infty}^{\infty} d x \left(\frac{\gamma}{2} \left[\frac{d h(x)}{d x} \right]^2 + \int\limits_0^{h(x)} dy \left[\rho g_e y - P -  p_c(x,y)\right] \right),
\end{equation}
and the pressure imbalance
$p_e(x)$ experienced by the interface at each site $x$ is given by $p_e(x) = - {\delta \mathcal H}/{\delta h(x)}$. The condition of mechanical equilibrium in Eq.~\eqref{eq:press_balance} corresponds therefore to setting $p_e(x) = 0$ \cite{HoltzmanCommPhys2020}.
In the framework of this model,
%applied pressure vs wetting-phase saturation trajectories (PS) 
PS trajectories
are built from the sequence of equilibria. 
The passage from one equilibrium configuration to the next 
%upon an infinitesimal change of pressure 
can be of two kinds.
In the first, the system remains trapped in a \emph{local} energy minimum, where the small change of external forcing $P$ causes a correspondingly small evolution of the wetting-phase saturation $S_w$, resulting in a smooth PS trajectory.
In the second, an abrupt change of state ($S_w$) takes place at the new value of $P$, in a Haines jump~\cite{Haines1930, Berg2013}; this occurs when the change of $P$ suppresses the current local energy minimum, and the system is forced to abruptly jump to a new metastable equilibrium.
Haines jumps are effectively instantaneous in the time scale of change of the driving pressure, so that interfacial configurations experience irreversible changes at punctuated values of $P$.

Numerically, this dynamics can be simulated iteratively by \emph{synchronous} updates of $h(x)$ in all unstable sites by a small amount in the direction that reduces $|p_e(x)|$, stopping when all sites retain equilibrium (for details of the numerical procedure, see \cite{HoltzmanCommPhys2020}). 
This deterministic rule is akin to the zero-temperature limit of the Glauber dynamics for RFIM \cite{SethnaPRL1993,Sethna2004}, which considers energy barriers between consecutive equilibria that are much larger than thermal fluctuations. 
%, considering that the energy barriers between consecutive metastable equilibria are much larger. It corresponds to the zero-temperature limit of the Glauber dynamics for random field Ising models \cite{SethnaPRL1993,Sethna2004}. 
%
The presence of the quenched disorder term $p_c[x,h(x)]$ in Eqs.~\eqref{eq:press_balance} and ~\eqref{eq:hamiltonian} defines a rugged free energy landscape, so that for every applied pressure there are many different interfacial configurations $h(x)$ that are local minimizers of $\cal{H}$. 
The synchronous dynamics described above takes the current configuration to the closest available metastable minimum in a deterministic manner dictated by the quenched disorder. 
Ref.\ \cite{HoltzmanCommPhys2020} proved that no parts of the interface recede under this dynamics, and a no-passing rule \cite{Middleton1992} is obeyed such that a configuration of larger (or equal) saturation compared to another will {remain so} under a monotonous evolution of the driving pressure. 
As a result, the original two-phase configuration is exactly recovered in any cyclic excursion of the driving pressure, a property known as RPM (return-point memory), ubiquitous to many athermal driven disordered systems \cite{Bertotti_V1,Goicoechea1994,KeimRMP2019,Bense2021}.

\subsection{Energy dissipated 
%during an interfacial  jump 
between equilibrium configurations}
\label{background_diss_equil}

The amount of energy dissipated can be obtained from the change in interfacial energy due to fluid displacement and the mechanical work done by the applied pressure $P$. For a small interface displacement $\delta h(x)$ the dissipated energy is
\begin{equation}
\label{eq:dPsi}
\dbar \Psi = dU - \dbar W,
\end{equation}
where $dU$ is the change in the internal energy, and
\begin{equation}
\dbar W=PdS=P\int_{-\infty}^{\infty}\delta h(x) dx
\label{energy_basic}
\end{equation}
is the work.
Here the notation $d\Box$ is used for infinitesimal changes of variables that are state functions (e.g. $S$), while $\dbar\Box$ is for changes of variables that are not (e.g. $W$). 
In our 2-D model, all energy units [Eq.~\eqref{energy_basic} and throughout] are of energy per unit length, i.e. the out-of-plane thickness.
$U$ accounts for the capillary energy of the front deformation
and for the gravitational potential energy of the oil phase (wetting fluid). We follow the convention
that $\dbar \Psi \le 0$.

Noting that $\mathcal H = U - PS$, we get $\delta \mathcal H = \dbar \Psi - S \, dP$. The fact that ${\mathcal H} = {\mathcal H} \left[ h(x), P \right]$ and $p_e = - \delta {\cal H} / \delta h(x)$ allows writing 
\begin{equation}
 \label{ch_hamilt}
 \delta {\mathcal H} = - \int_{-\infty}^{\infty} d x \, p_e(x) \, \delta h(x) - \int_{-\infty}^{\infty} d x \int_0^{h(x)} d y \, dP.
\end{equation}
From this, as well as from Eq.~\eqref{eq:hamiltonian} and the expression ${\cal H}=U-PS$, we can compute the internal energy as 
\begin{equation}
\label{eq:internal_energy}
U = \int\limits_{-\infty}^{\infty} d x \left[\frac{\gamma}{2} \left(\frac{\partial h}{\partial x} \right)^2 + \int\limits_0^{h} dy \left(\rho g_e y - p_c\right) \right]
\end{equation}
Finally, the energy lost between two equilibrium states, $t-1$ and $t$, is found by integrating Eq.~\eqref{eq:dPsi}, which gives~\cite{Holtzman2023}
\begin{equation}
\label{eq:dissipation}
\Psi^{t-1 \to t} = \left[ U^t - U^{t-1} \right] 
- P^t \left[ S_w^t - S_w^{t-1} \right].
\end{equation}

\subsection{Energy dissipated per incremental change in interface configurations}

Here, we present an alternative method for computing energy dissipation between consecutive interfacial configurations.
For continuous, reversible displacements (isons), where the system stays in one local minimum of $\mathcal H$, and a small change in $P$ leads to a small change in $h(x)$, $p_e(x)=0$ by definition. This reduces Eq.\ (\ref{ch_hamilt}) to $\delta {\mathcal H} = - \int_{-\infty}^{\infty} d x \int_0^{h(x)} d y \, dP = - S \, dP,$
which proves that $\dbar \Psi = 0$, i.e., no dissipation. Since there are no energy losses in this case, $\dbar W=dU$, and for a finite continuous displacement $W = \Delta U$, with $\Delta U$ 
%$\int \dbar W=\int dU$, 
%thus, the total work $\int PdS$ must be identical to the energy change 
computed from Eq.~\eqref{eq:internal_energy}.

In irreversible displacements (rheons) $P$ is constant ($dP = 0$) and $\delta \mathcal H =  \dbar \Psi$. This in turn provides the energy dissipated for each elementary step $\delta h$ within a Haines jump, using Eq.\ (\ref{ch_hamilt}):
\begin{equation}
\label{diff_dissip_irreversible}
\dbar \Psi = \delta {\mathcal H} = - \int_{-\infty}^{\infty} d x \, p_e(x) \, \delta h(x).
\end{equation}
To obtain the total energy dissipated in a given jump between equilibrium states $t-1$ and $t$, $\Psi^{t-1 \to t}$, one can integrate Eq.~\eqref{diff_dissip_irreversible} over all the intermediate nonequilibrium steps $k$,
\begin{equation}
\label{total_dissip_irreversible}
\Psi^{t-1 \to t}= - \sum_{k} \int_{-\infty}^{\infty} d x \, p_e^k(x) \, \delta h^k(x).
\end{equation}
Equations~\eqref{diff_dissip_irreversible} and \eqref{total_dissip_irreversible} present the energy dissipated as the sum of elementary products of unbalanced pressure $p_e(x)$ times the corresponding displacement $\delta h(x)$. 
This will be used below (Section \ref{sec:config_single}) to provide an explanation for a nonintuitive observation: asymmetry between energy dissipation in drainage and imbibition.
Furthermore, Eqs.~\eqref{diff_dissip_irreversible}--\eqref{total_dissip_irreversible} show that the energy dissipation within avalanches does not have to be proportional to the corresponding change in saturation, as the values of $p_e(x)$ can be different from site to site. This non-proportionality between changes in saturation and dissipation was shown numerically in disordered media constructed from defects of various strengths \cite{Holtzman2023}. A related finding was shown for quasistatically driven disordered ferromagnets \cite{Ortin1998}. 
An extreme example of this non-proportionality arises in the limit in which dissipation approaches zero even as the size of the interface jump remains finite, {as we will see below}.

%%%%%%%%%%%%%%%%%%%%%%%%%%%%%%%%%%%%%%%%%%%%%%%%
% \section{Single defect system}
\section{Interface configurations: Single defect}
\label{sec:config_single}
%%%%%%%%%%%%%%%%%%%%%%%%%%%%%%%%%%%%%%%%%%%%%%%%%

\begin{figure}%[h]
\includegraphics[width=.45\textwidth]{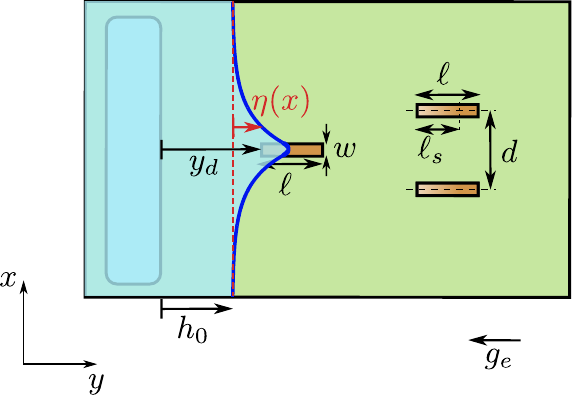}
		\centering
 \caption{
Our 2-D model system representing an imperfect Hele-Shaw cell with local perturbations in thickness {of width $w$ and length $\ell$} (defects; in brown), which alter the out-of-plane capillary pressure $p_c$. 
Once the fluid-fluid interface (blue line) passes through a defect, the interface deforms by $\eta(x)$, measured relative to the flanks (which are at $y=h_0$) such that the interface height $h(x)=h_0+\eta(x)$ is nonuniform.
We consider both ``mesa'' defects (fixed $p_c$ {within the defect}) and regular defects which include a sloping part of length $\ell_s$ (linear increase in $p_c$) and a plateau (fixed $p_c$).  
We also consider both a single, isolated defect, and a pair of interacting defects (separated by $d$); for simplicity, the interface, as well as its descriptors [$h_0$, $\eta(x)$], and the defect position $y_d$ are not shown for the defect pair. 
 }
\label{Fig1_sketch_setup}
 \end{figure}

In this Section, we formulate an analytical solution for the equilibrium interface configuration for a single defect. 
When an incompressible wetting fluid (e.g. oil) imbibes or drains quasistatically in a smooth Hele-Shaw cell with a narrow gap of fixed width, filled with inviscid, non-wetting fluid (e.g. air), the interface is morphologically stable and $h(x)=h_0$. Modulations of the out-of-plane capillary pressure, $p_c(x,y) = p_c^{0} + \delta p_c(x,y)$, cause the interface to deform \cite{DiazPiola2017,Planet2020,HoltzmanCommPhys2020}. 
Expressing the modulation as $\delta p_c(x,y)=\delta p_c^* F(x,y)$, where 
$\delta p_c^*$ is the maximum value and
$F(x,y)$ is the ``normalized modulation'', 
we can rewrite Eq.~\eqref{eq:press_balance} as~\cite{Planet2020}
\begin{equation}
\label{eq:pressure_balance_2}
\gamma \frac{d^2 \eta(x)}{d x^2} - \rho g_e \eta(x)  + \delta p_c^* F[x,\eta(x)+h_0]  = 0\, ,
\end{equation}
where $\eta(x) = h(x) - h_0$ (Fig~\ref{Fig1_sketch_setup}).
Note that $h_0 = (P + p_c^0)/\rho g_e$.
In the following we consider the case of $\delta p_c^* > 0$ ($\delta p_c^* < 0$ is analogous). 
{We consider a defect with $F(x,y)$ nonzero within a rectangle $-w/2<x<w/2$, $y_d<y<y_d+\ell$, where $w$ is the width of the defect and $\ell$ its length. 
We do not set $y_d = 0$ to keep our formulation general such that it can be used for a disordered system with multiple defects.}

%
%%%%%%%%%%%%%%%%%%%%%%%%%%%%%%%%%%%%%
%%%%%%%%%%%%%%%%%%%%%%%%%%%%%%%%%%%%%

The effective pressure field given by the left-hand side of Eq.~\eqref{eq:pressure_balance_2} can be split into two parts: $p_e(x)=p_d(x)+\delta p_c^* F[x,\eta(x)+h_0]$, where $p_d$ accounts for the restoring force of the line and it is linear in $\eta$~\cite{Planet2020}.
We obtain the equilibrium states, $p_e = 0$, by equating
\begin{equation}
    p_d(x)=-\delta p_c^* F[x,\eta(x)+h_0].
    \label{eq:pdpc}
\end{equation}
Here, we find two different scenarios depending on the number of possible equilibrium configurations (roots) $\eta_c$ that fulfill Eq.~\eqref{eq:pdpc}.
If for all $h_0$ there is only one root, the defect does not lead to hysteresis or dissipation and is termed ``weak''. 
In contrast, for an hysteretic and dissipative (``strong'') defect, for some interval of $h_0$ the solution gives three equilibrium points: two stable equilibrium configurations, with the largest and smallest $\eta$, while the intermediate one is unstable. 
We note that for complex $p_c$ profiles it is possible to have more than three roots; here we consider only linear variations in $p_c$.

% shows sketches of 
% regular and non-regular defects.
%
\begin{figure}
\includegraphics[width=.32\textwidth]{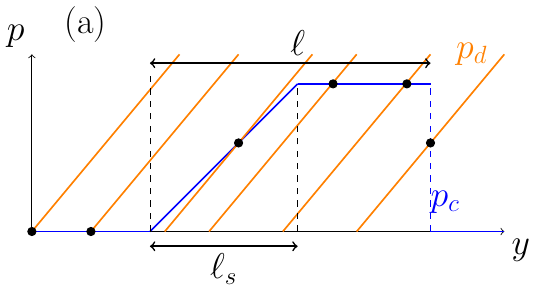}
\includegraphics[width=.32\textwidth]{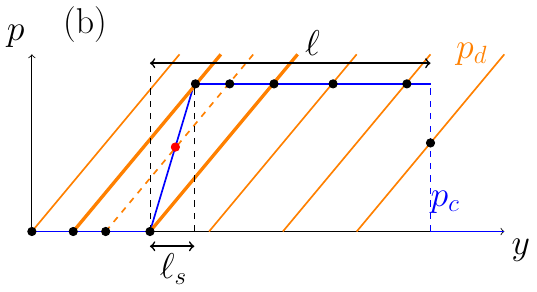}
\includegraphics[width=.32\textwidth]{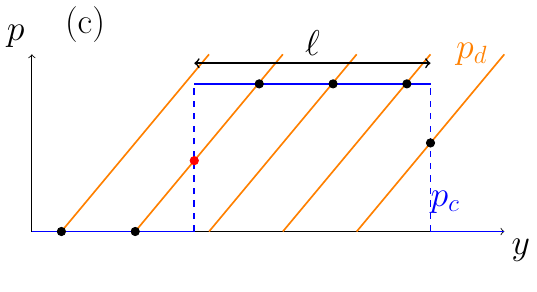}
		\centering
 \caption{Geometrical construction of the equilibrium positions of the interfaces in the presence of a localized defect:
 % \hl{Explicar que es cada cosa que es veu a la figura}. 
 (a) weak; (b) strong; and (c) mesa.
 Blue lines correspond to the pressure field exerted by the defect. 
 Orange lines correspond to the restoring force (pressure) of the interface, opposing the deformation.
 At equilibrium, $p_d = p_c$, and these states are presented as black dots for stable equilibrium.
 Red dots correspond to unstable states.
}
\label{Fig2:pressure_balance}
 \end{figure}

Figure~\ref{Fig2:pressure_balance} shows visual representations of these different scenarios. 
{These 1-D representations are exact for infinitely wide defects (when there is no $x$ dependence), and remain qualitatively valid for defects of arbitrary width.} 
We refer to panels (a--b) as ``regular'' defects, where the change of $p_c$ is continuous along the $y$ axis. Panel (c) shows the special case of the ``mesa'' defect \cite{Joanny1984}, where the modulation $F[x,\eta(x)]$ is a rectangular function in $y$, so that the change in $p_c$ is discontinuous (non-regular).
The 2-D interface shape when passing through a gap modulation can be derived by realizing that Eq.~\eqref{eq:pressure_balance_2}
can be written as $\delta p_c^* F[x,\eta(x)+h_0] = -\mathcal{L}\eta(x)$ with a linear differential operator $\mathcal{L}$. The interface shape $h(x)$ is obtained using the Green's function formalism \cite{Planet2020}
\begin{equation}
\label{eq:hG}
\eta(x) = \int\limits_{-\infty}^\infty dx' G_0(x-x') \delta p_c^* F[x',h_0+\eta(x')],
\end{equation}
where $G_0(x)$ satisfies 
\begin{equation}
\gamma \frac{d^2 G_0(x)}{d x^2} - \rho g_e G_0(x) = \delta(x)\, ,
\end{equation}
It takes the form~\cite{Planet2020}
\begin{equation}
\label{eq:G0}
G_0(x) = \frac{\ell_c}{2 \gamma} \exp(-|x|/\ell_c),
\end{equation}
with $\ell_c = \sqrt{\gamma/\left(\rho g_e\right)}$.

\subsection{Interface configurations: Mesa defect}

For the special case of a 
%an irregular 
mesa defect,
\begin{equation}
    F(x,y) =\displaystyle {\rm Rect}\left(\frac{x}{w}\right) {\rm Rect}\left(\frac{y-(y_d+\ell/2)}{\ell}\right)\, ,
\end{equation}
where ${\rm Rect}(x)$ is the rectangular function ${\rm Rect}(x)=H(x+1/2)-H(x-1/2)$, $H(x)$ is the Heaviside function, $w$ is the width and $\ell$ the length of the defect. When passing such defects, the interface displacement is {\it always} hysteretic. 
The analytical solutions for the interface shape when passing through a mesa defect, derived in  \cite{HoltzmanCommPhys2020}, 
are summarized below for completeness.

Assuming that for $|x|<w/2$, $h_0+\eta$ is between $y_d$ and $y_d+\ell$, the integral in Eq.~\eqref{eq:hG} can be evaluated analytically, which gives
\begin{equation}
\label{eq:mesa_solution}
\eta(x) = \frac{\delta p_c^*}{\rho g_e}\begin{cases}
\exp(-|x|/\ell_c) \sinh(w/2\ell_c)\, , & |x| > w/2\\
1 - \exp(-w/2 \ell_c) \cosh(x/\ell_c)\, , & |x| \leq w/2
\end{cases}
\end{equation}
The maximum deformation occurs at the defect centre $x = 0$, and equals
\begin{equation}
\eta_m = \frac{\delta p_c^*}{\rho g_e}\left[1 - \exp(-w/2 \ell_c)\right]\, .
\end{equation}
During imbibition, the interface remains flat until it contacts the defect, $h_0=y_d$, at which point it deforms abruptly to a shape given by Eq.~\eqref{eq:mesa_solution}; here and elsewhere in this paper, we assume that the defect is sufficiently long to accommodate the deformed interface, i.e., $\ell>\eta_m$. 
At the end of the defect, $y=y_d+\ell$, the interface gets pinned and deformation decreases continuously to zero (flat) by {the time} $h_0=y_d+\ell$, at which point the interface exits the defect.
During drainage, the initial part of the interface displacement is reversible (identical to that in imbibition),
%moves reversibly from the position above the defect towards the defect, where it 
i.e. it gets pinned and deforms
until $h_0 = (y_d + \ell) - (\delta p_c^*/\rho g_e) \left[1 - \exp(-w/2 \ell_c)\right]$. 
%This evolution for decreasing $h_0$ is fully reversible. 
However, when $h_0$ reaches $y_d$ (the point of interface jump in imbibition), the trivial solution $\eta=0$ appears as a metastable solution, but {the interface} remains deformed (thus, hysteresis is observed). 
The jump {is delayed until} $h_0$ is equal to the external head $h_c$, given by
\begin{equation}
\label{eq:hc}
h_c = y_d - \frac{\delta p_c^*}{2 \rho g_e} \left[1 -\exp(-w/\ell_c)\right]\,. 
\end{equation}
At that point the interface passes through the corners of the defect. For lower $h_0$, the effective width $w_e$ across which the defect is wet would have to be smaller than $w$, and the new equilibrium position would be
\begin{equation}
h_0 = y_d - \frac{\delta p_c^*}{2 \rho g_e} \left[1 -\exp(-w_e/\ell_c)\right] > h_c. 
\end{equation}
However, this is not possible~\cite{HoltzmanCommPhys2020} and the nontrivial solution disappears.

Note that for a very wide defect ($w\gg\ell_c$), {immediately before the jump during drainage} the interface profile around $x=-w/2$ can be approximated as
\begin{equation}
h(x)={y_d+}\frac{\delta p_c^*}{2\rho g_e}\begin{cases}
\exp(\Delta x/\ell_c)-1\ , & \Delta x<0,\\
1-\exp(-\Delta x/\ell_c)\ , & \Delta x>0.
\end{cases}
\end{equation}
where $\Delta x=x+w/2$ and $|\Delta x|\ll w$. Thus, the solution is approximately symmetric with respect to the corner of the defect ($x=-w/2$, $y=y_d$) as the center of symmetry [$h(-w/2+\Delta x)-y_d\approx -\{h(-w/2-\Delta x)-y_d\}$]. 
    %\hl{Is this clear?} 
Similarly for $x=w/2$ and the corresponding corner. We will compare this to the case of a wide regular defect with a slope below.

\subsection{Interface configurations: Regular defect}

We now consider a defect with a capillary pressure profile which varies in $y$,
\begin{equation}
\label{pcxz}
F(x,y) = a_1(x)c(y),
\end{equation}
where
\begin{equation}
\label{eq:ax}
a_1(x)={\rm Rect}\left(\frac{x}{w}\right)
\end{equation}
and
\begin{equation}
\label{eq:cy}
c(y)=\frac{1}{\ell_s}(y - y_d) {\rm Rect} \left(\frac{y-y_d-\ell_s/2}{\ell_s}\right) + {\rm Rect} \left(\frac{y-y_d-(\ell_s+\ell)/2}{\ell-\ell_s}\right).
\end{equation}
The defect has width $w$ and length $\ell$, with the capillary pressure profile consisting of two parts: a slope (ramp) of length $\ell_s$ where $p_c$ increases linearly in $y$ and a plateau of fixed $p_c$ (as in the mesa defect, cf. Fig.~\ref{Fig1_sketch_setup}).
In the ramp, the pressure slope is
\begin{equation}
\Pi=\frac{dp_c(0,y)}{dy}=\frac{\delta p_c^*}{\ell_s}.
\end{equation}
The interface deformation is calculated by substituting Eq.~\eqref{pcxz} in the integral equation~\eqref{eq:hG}, 
\begin{equation}
\label{IE0}
\eta(x) = \delta p_c^* \int\limits_{-\infty}^{\infty} dx' G_0(x-x') a_1(x')c[h_0+\eta(x')],
\end{equation}
where the 
%\hl{unperturbed} 
propagator $G_0(x)$ is given by Eq.~\eqref{eq:G0}.

In imbibition, prior to touching the defect, $h_0<y_d$, the interface is undeformed, $\eta(x)=0$. 
As the interface enters the defect, $h_0>y_d$, as {long as the interface deformation is small enough such that {inside the defect} it remains within the ramp}, 
Eq.~\eqref{IE0} becomes
\begin{equation}
\label{IE}
\eta(x)=\Pi \int\limits_{-\infty}^{\infty} dx' G_0(x-x')a_1(x')(h_0-y_d)+\Pi \int\limits_{-\infty}^{\infty} dx' G_0(x-x')a_1(x')\eta(x').
\end{equation}
We distinguish between two cases: the defect is \emph{weak} if a solution of this equation exists, {at least for small enough positive $h_0-y_d$}; it then changes continuously as $h_0$ increases.
The defect is \emph{strong} if there are no solutions {for positive $h_0-y_d$}; in that case, upon entering the defect the interface experiences an abrupt jump that extends into the plateau.

To obtain a closed-form analytical solution, we consider below two limits:  infinitely narrow and infinitely wide defects. 
For these, we find the threshold between weak and strong, and the interface shapes before and after the jump in the strong regime. 
For the general case of a finite defect width, the threshold calculation is provided in 
%The threshold for the general case is obtained in 
Appendix~\ref{appendix_critical_slope}, and the numerical methods for establishing the interface shape are described in Appendix~\ref{appendix_profiles}.

The threshold between weak and strong {defects} in terms of the pressure slope, $\Pi_c^{(1)}$, is independent of $\ell_s$ given $\ell_c$, $w$ and $\gamma$. Dimensionality considerations provide
\begin{equation}
\label{eq:pc1gen}
\Pi_c^{(1)}=\frac{\gamma}{\ell_c^2}f\left(\frac{w}{\ell_c}\right),
\end{equation}
where $f$ is a dimensionless function. Given $\Pi/\Pi_c^{(1)}$ and $w/\ell_c$, the interface deformations before the jump in imbibition, $\eta_{\rm imb}$, and after the jump in drainage, $\eta_{\rm dra}$, are proportional to $\ell_s$, thus,
\begin{equation}
\label{eq:etaimbgen}
\eta_{\rm imb}(x)=\ell_s \phi_{\rm imb}\left(\frac{x}{\ell_c};\frac{\Pi}{\Pi_c^{(1)}},\frac{w}{\ell_c}\right)
\end{equation}
and
\begin{equation}
\label{eq:etadragen}
\eta_{\rm dra}(x)=\ell_s \phi_{\rm dra}\left(\frac{x}{\ell_c};\frac{\Pi}{\Pi_c^{(1)}},\frac{w}{\ell_c}\right),
\end{equation}
where $\phi_{\rm imb}$ and $\phi_{\rm dra}$ are dimensionless functions.

%%%%%%%%%%%%%%%%%%%%%%%%%%%%%%%%%%%%%%%%%
\subsubsection{Narrow defect limit ($w \ll \ell_c$)}
%%%%%%%%%%%%%%%%%%%%%%%%%%%%%%%%%%%%%%%%%
\label{sssec:system_narrow}
Consider a narrow defect, $w \ll \ell_c$, located at $x = 0$. We note that $a_1(x) = w \delta_w(x)$
with $\delta_w(x) = w^{-1} {\rm Rect} (x/w)$, where in the limit $w \to 0$, $\delta_w(x)$ approaches the Dirac delta $\delta(x)$. 
% Using expression~\eqref{a1} for $\delta_w(x) \to \delta(x)$ in Eq.~\eqref{IE}, we obtain
Introducing this approximation in Eq.~\eqref{IE}, we obtain
\begin{equation}
\eta(x) = \Pi w (h_0 - y_d) G_0(x) +  \Pi w \eta_0 G_0(x), 
\end{equation}
where $\eta_0=\eta(0)$. For $x=0$, this becomes
\begin{equation}
\eta_0 = \Pi w (h_0 - y_d) \beta_0 +  \Pi w \eta_0 \beta_0,
\end{equation}
with $\beta_0 = G_0(0) = \ell_c/2\gamma$. This gives 
    %\hl{RH: use of $\alpha_0$ and $\alpha_1$ may be confusing with our use of $\alpha$ as tilt angle?} {\color{red}MC: I've replaced by $\beta$}
%
\begin{equation}
\eta_0 = \frac{\Pi w \beta_0 (h_0 - y_d)}{1 - \Pi w \beta_0},
\end{equation}
so that we obtain the following solution:
\begin{equation}
\label{eq:soleta}
\eta(x) = G_0(x) \frac{\Pi w (h_0 - y_d)}{1 - \Pi w \beta_0}.
\end{equation}
This is consistent with our assumption that the solution for the interface configuration {crosses the defect} within the ramp when $y_d < h_0 + \eta_0 < y_d + \ell_s$. This condition is satisfied for sufficiently small and positive $h_0-y_d$, when the denominator of Eq.~\eqref{eq:soleta} is positive, i.e.,
\begin{equation}
\Pi<\Pi_n^{(1)}=\frac{1}{w\beta_0}=\frac{2\gamma}{w\ell_c}.
\label{eq:crit_slope_weak_single}
\end{equation}
In this case, the defect is weak. 
Here $\Pi_n^{(1)}$ is the narrow-defect approximation for the critical slope in a single defect, where the general threshold for a single defect, $\Pi_c^{(1)}$, is derived in Appendix~\ref{appendix_critical_slope}. 
Conversely, for $\Pi>\Pi_n^{(1)}$ there is no solution {crossing the defect} within the ramp for positive $h_0-y_d$, indicating that the interface deforms abruptly beyond the sloping part, and the defect is strong. Note that a solution {crossing the defect} within the ramp does exist for strong defects when $h_0-y_d$ is small enough 
%by absolute value 
and \textit{negative}; this is the unstable solution marked by the red dot in Fig.~\ref{Fig2:pressure_balance}(b).

For the case where the interface deforms beyond the {ramp} length and reaches the plateau region, Eq.~\eqref{IE0} gives
\begin{equation}
\eta(x)=\delta p_c^* w G_0(x)=\delta p_c^* \frac{\ell_c w}{2\gamma}\exp(-|x|/\ell_c)=\frac{\Pi}{\Pi_n^{(1)}}\ell_s\exp(-|x|/\ell_c).
\end{equation}
In cases where the defect width cannot be neglected, we make the assumption that the deformation is constant within the defect, providing
\begin{equation}
\label{eq:narrow_plateau}
\eta(x)=\frac{\Pi}{\Pi_n^{(1)}}\ell_s\begin{cases}
\exp[-(|x|-w/2)/\ell_c], & |x|>w/2,\\
1, & |x|<w/2.
\end{cases}
\end{equation}
As expected, this coincides with the $w\ll\ell_c$ limit of the mesa defect case, Eq.~\eqref{eq:mesa_solution}. 
The deformation is independent of $h_0$ as long as the interface solution crosses the defect within the plateau. 
For a strong defect ($\Pi/\Pi_n^{(1)}>1$), this occurs {already} for $h_0=y_d$ 
%(we assume here and in what follows that the defect is long enough, i.e., 
[again, considering a sufficiently long defect, $\eta(0)=(\Pi/\Pi_n^{(1)})\ell_s<\ell$].
%\hl{Repetition but I think we keep, see above}.
Thus, during imbibition, the interface jumps from
\begin{equation}
\label{eq:narrow_imbm}
h=h_{\rm imb}^{-}=y_d
\end{equation}
to
\begin{equation}
\label{eq:narrow_imbp}
h=h_{\rm imb}^{+}=y_d+\frac{\Pi}{\Pi_n^{(1)}}\ell_s\begin{cases}
\exp[-(|x|-w/2)/\ell_c], & |x|>w/2,\\
1, & |x|<w/2.
\end{cases}
\end{equation}

In drainage, in the beginning the interface displacement is identical to the mesa case, i.e., it pins at the edge of the defect, $y=y_d+\ell$, until the deformation reaches the value given by Eq.~\eqref{eq:narrow_plateau}, and then moves continuously. 
The jump in drainage occurs when the interface reaches the boundary between the plateau and the ramp (rather than the end of the defect, $y=y_d$, in the mesa case), and the solution for the interface configuration, \eqref{eq:narrow_plateau}, meets the unstable branch and ceases to exist. 
At this point, $h_0={y_d+}\ell_s-\eta(0)={y_d}-\ell_s(\Pi/\Pi_n^{(1)}-1)$, and the jump occurs between
\begin{equation}
\label{eq:narrow_dram}
h_{\rm dra}^{-}=\begin{cases}
y_d-\ell_s\left\{{\displaystyle\frac{\Pi}{\Pi_n^{(1)}}}[1-\exp(-(|x|-w/2)/\ell_c)]-1\right\}, & |x|>w/2,\\
y_d+\ell_s, & |x|<w/2
\end{cases}
\end{equation}
and
\begin{equation}
\label{eq:narrow_drap}
h_{\rm dra}^{+}=y_d-\ell_s(\Pi/\Pi_n^{(1)}-1).
\end{equation}

\begin{figure}%[h]
\includegraphics[width=.45\textwidth]{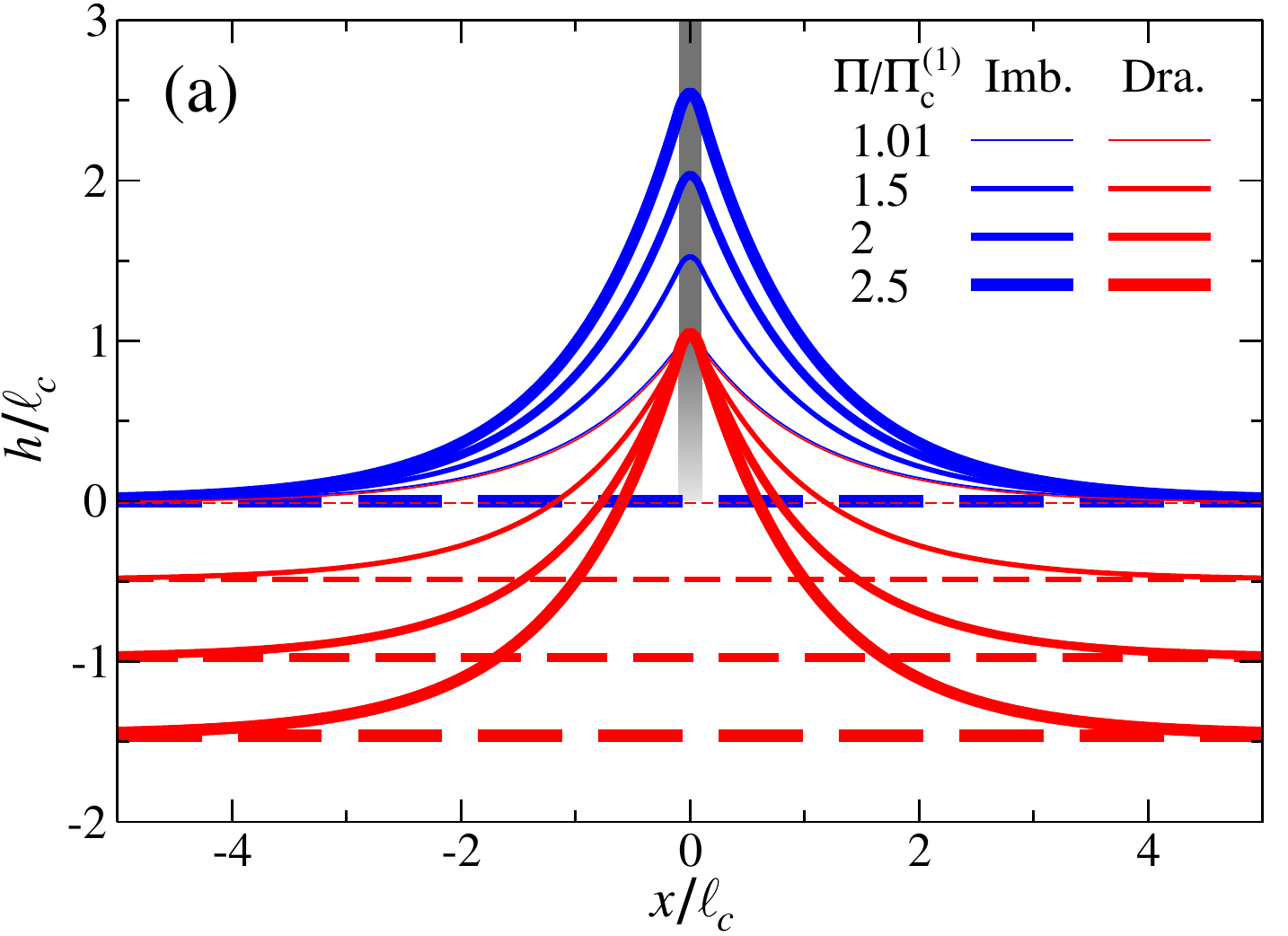}
\hspace{0.2cm}
\includegraphics[width=.45\textwidth]{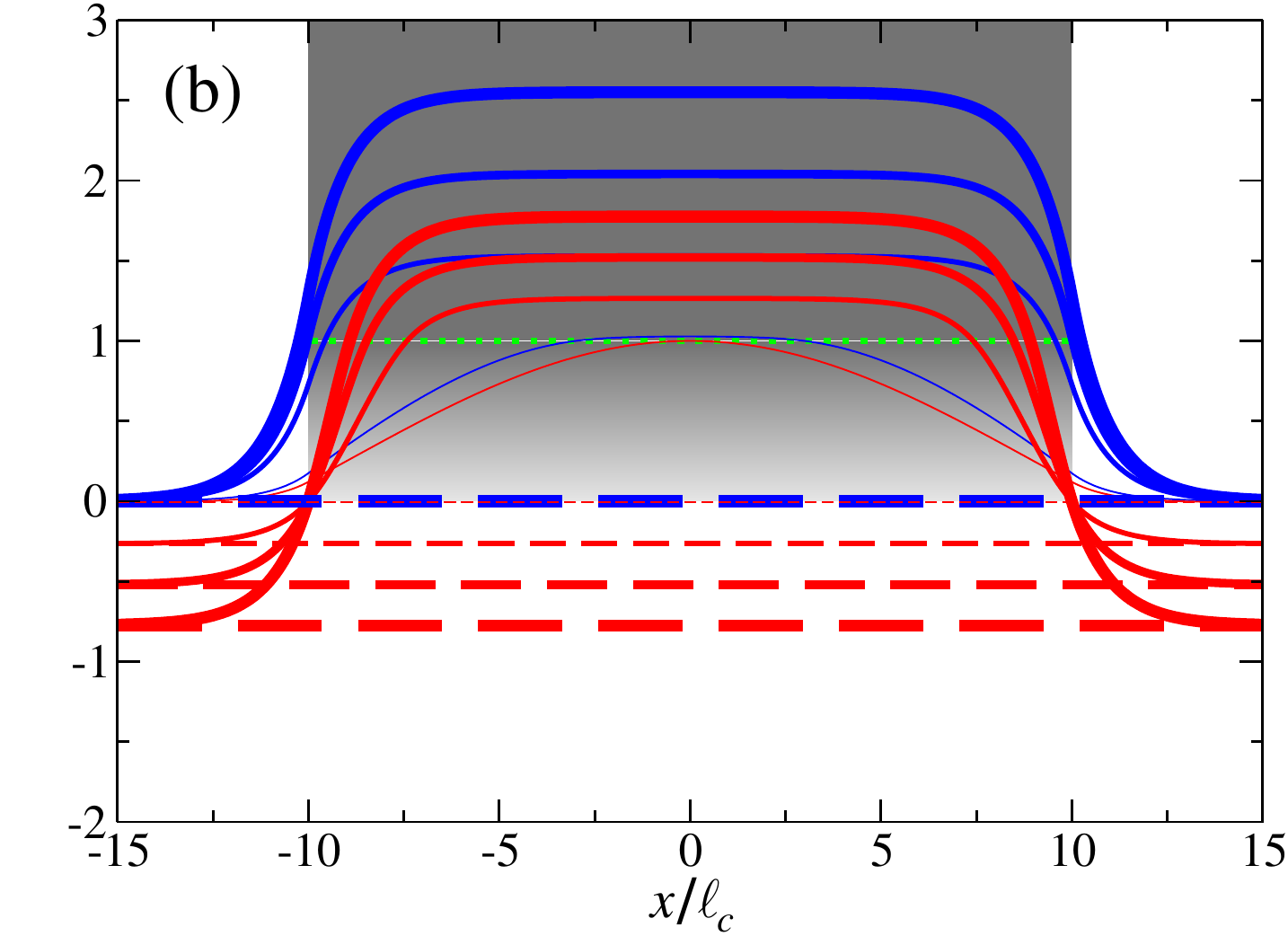}	\centering
 \caption{
     Interface profiles before and after jumps for a narrow [$w/\ell_c=0.2$; (a)] and a wide [$w/\ell_c=20$; (b)] regular ramp defects, for different capillary pressure slopes. The ramp part of the defect is shown by a shading gradient, and the plateau part in uniform gray. The boundary between these two regions is at $h/\ell_c=1$ and in panel (b) is marked with a green dotted line. Dashed and solid lines represent the undeformed (flat) and deformed configurations, i.e. before and after the jump in imbibition, and the opposite in drainage.
}
\label{Fig:profiles}
\end{figure}

These interfacial jumps at various defect strengths, computed numerically for a sufficiently narrow defect $w/\ell_c=0.2$ (Appendix~\ref{appendix_profiles}), are illustrated in Fig.~\ref{Fig:profiles}a.
As the transition between weak and strong is approached, $\Pi/\Pi_n^{(1)}\to 1+$, the difference between $h_{\rm imb}^{+}$ and $h_{\rm dra}^{-}$ vanishes (barely noticeable for $\Pi/\Pi_n^{(1)}=1.01$), although the jump remains finite (noting it can vanish for other defect profiles, not considered here).

%%%%%%%%%%%%%%%%%%%%%%%%%%%%%%%%%%%%%
%%%%%%%%%%%%%%%%%%%%%%%%%%%%%%%%%%%%%

%%%%%%%%%%%%%%%%%%%%%%%%%%%%%%%%
\subsubsection{Wide defects ($w\gg\ell_c$)}
\label{ssec:infidef}
%%%%%%%%%%%%%%%%%%%%%%%%%%%%%%%%
In case of a very wide defect, the width of the Green's function $G_0$ [which is on the order of $\ell_c$; see Eq.~\eqref{eq:G0}] is much smaller than the width of the defect, and $G_0$ can be approximated by a delta function,
\begin{equation}
G_0(x)\approx\frac{\ell_c^2}{\gamma}\delta(x),
\end{equation}
such that Eq.~\eqref{eq:hG} becomes
\begin{equation}
\eta(x)=\frac{\ell_c^2}{\gamma}\delta p_c^* F[x,h_0+\eta(x)].
\end{equation}
This approximation is valid far from lateral boundaries of the defect. Outside the defect, $\eta(x)=0$. If the interface solution {crosses the defect} within the ramp, then inside the defect we get from the above 
\begin{equation}
\eta=\frac{\ell_c^2 \Pi}{\gamma}(h_0+\eta-y_d),
\end{equation}
{and from this}, or, equivalently from Eq.~\eqref{IE} with the same approximation for $G_0$,
\begin{equation}
\label{eq:solinf}
\eta=\frac{(\ell_c^2\Pi/\gamma)(h_0-y_d)}{1-\ell_c^2\Pi/\gamma}=\frac{\Pi(h_0-y_d)}{\rho g_e-\Pi}.
\end{equation}
This solution {crosses the defect} within the ramp for small enough positive $h_0-y_d$ if
%\begin{equation}
$\Pi<\Pi_w^{(1)}=\rho g_e={\gamma} / {\ell_c^2}$.
%\end{equation}
In this case, the defect is weak. 
Here $\Pi_w^{(1)}$ is the wide-defect approximation for the general solution for the threshold for a single defect, $\Pi_c^{(1)}$, derived in Appendix~\ref{appendix_critical_slope}. Otherwise (for {$\Pi>\Pi_w^{(1)}$}), during imbibition the interface jumps into the plateau upon touching it at $y_d$. In that case, the deformation of the interface part which is inside the defect {not too close to its edges} is $\eta=\delta p_c^*/\rho g_e=\ell_s(\Pi/\Pi_w^{(1)})$, similar to that for the mesa defect [namely Eq.~\eqref{eq:mesa_solution} for $|x| < w/2$ except near the defect edges].
Thus, the interface configuration before and after the jump is
\begin{equation}
\label{eq:wide_imbm}
h_{\rm imb}^{-}=y_d
\end{equation}
and
\begin{equation}
\label{eq:wide_imbp}
h_{\rm imb}^{+}=\begin{cases}
y_d, & |x|>w/2,\\
y_d+\ell_s(\Pi/\Pi_w^{(1)}), & |x|<w/2.
\end{cases}
\end{equation}
%This analytical expression is confirmed by the numerical computations (Appendix~\ref{appendix_profiles}), see Fig.~\ref{Fig:profiles}b. 
An exception to the validity of this calculation is for $\Pi/\Pi_w^{(1)}$ just above and very close to 1. In such case, $\Pi$ may be above $\Pi_w^{(1)}$ yet below the exact threshold $\Pi_c^{(1)}$; even if not, the influence of the edges extends very far inside the defect (see Fig.~\ref{Fig:profiles}b). 
For drainage (except for $\Pi/\Pi_c^{(1)} \approx 1$)
immediately before the jump the interface passes {very close to} the corners of the defect at $y=y_d$; this is similar to the \emph{mesa} but not to the \emph{regular} narrow defect case. 
It can be shown then that the interface shape around the defect edge at $x=-w/2$ still has a center of symmetry, but at height $y_d+\ell_s/2$, instead of in the corner of the defect. Mathematically, $h(x_0+\Delta x)-y_d-\ell_s/2\approx -[h(x_0-\Delta x)-y_d-\ell_s/2]$ when $|\Delta x|\ll w$, where $x_0$ is such that $0<x_0+w/2\ll w$, and analogously for $x$ near $w/2$. 
    %\hl{OK?}
The interfacial configurations {before and after the jump} in drainage are {then}
\begin{equation}
\label{eq:wide_dram}
h_{\rm dra}^{-}=\begin{cases}
y_d-\ell_s{\displaystyle\frac{\Pi/\Pi_w^{(1)}-1}{2}}, & |x|>w/2,\\[0.3cm]
y_d+\ell_s{\displaystyle\frac{\Pi/\Pi_w^{(1)}+1}{2}}, & |x|<w/2.
\end{cases}
\end{equation}
and
\begin{equation}
\label{eq:wide_drap}
h_{\rm dra}^{+}=y_d-\ell_s\frac{\Pi/\Pi_w^{(1)}-1}{2}.
\end{equation}
These analytical expressions are confirmed by the numerical computations (Appendix~\ref{appendix_profiles}), see Fig.~\ref{Fig:profiles}b.              
%

%%%%%%%%%%%%%%%%%%%%%%%%%%%%%%%%%%%%%%%%%%%%%%%%%
%%%%%%%%%%%%%%%%%%%%%%%%%%%%%%%%%%%%%%%%%%%%%%%%%
%%%%%%%%%%%%%%%%%%%%%%%%%%%%%%%%%%%%%%%%%%%%%%%%%
% \section{Energy balance for %quasistatic displacements across 
% a single defect}
\section{Energy balance: Dissipation during jumps}
%Single defect}
\label{sec:energy_balance}
%%%%%%%%%%%%%%%%%%%%%%%%%%%%%%%%%%%%%%%%%%%%%%%%%
%%%%%%%%%%%%%%%%%%%%%%%%%%%%%%%%%%%%%%%%%%%%%%%%%
%%%%%%%%%%%%%%%%%%%%%%%%%%%%%%%%%%%%%%%%%%%%%%%%%

We begin by considering a closed imbibition-drainage cycle for a strong defect.
%Consider a cycle in which first the viscous fluid is forced to invade completely a strong defect, and after that the fluid is drained up to the original position.
%
The first dissipation event occurs when the interface enters the defect in imbibition. The external pressure is $\rho g_e h_0 = \rho g_e h_{\rm imb}^{-}=\rho g_e y_d$, and the work
% $H_e = y_d$
%$H_e = d$ 
is 
$ W_{\rm imb} = \rho g_e y_d \int\limits_{-\infty}^\infty d x\, \eta_{\rm imb}(x)$;
here we used the fact that the deformation after the jump, $\eta_{\rm imb}(x)$, equals $h_{\rm imb}^{+}-h_{\rm imb}^{-}$.
The internal energy change during this deformation is 
 \begin{align}
\Delta U_{\rm imb} &= U\{h_{\rm imb}^{+}\} - U\{h_{\rm imb}^{-}\} 
\nonumber\\
&= \int\limits_{-\infty}^{\infty} dx
  \left[\frac{\gamma}{2} \left(\frac{d \eta_{\rm imb}}{dx}\right)^2  + \frac{\rho g_e}{2} (\eta_{\rm imb}^2 + 2 y_d \eta_{\rm imb}) 
  - \delta p_c^* \int_{y_d}^{y_d + \eta_{\rm imb}} dy\, F(x,y)
  \right], 
\end{align} 
where $U\{h\}$ denotes the functional dependence of the internal energy on the interface configuration. Thus, we obtain for the dissipated energy 
\begin{equation}
\Psi_{\rm imb} =\Delta U_{\rm imb}-W_{\rm imb} = \int\limits_{-\infty}^{\infty} dx
  \left[\frac{\gamma}{2} \left(\frac{d \eta_{\rm imb}}{dx}\right)^2  + \frac{\rho g_e}{2} \eta_{\rm imb}^2  
  % - \eta \delta p_c^* \Theta(w/2-|x|)
  - \delta p_c^* \int_{y_d}^{y_d + \eta_{\rm imb}} dy\, F(x,y)
  \right].
  \label{eq:diss_dif}
\end{equation}

The second dissipation event occurs when the interface leaves the defect in drainage.
% (always considering the conditions set in the preamble, see Sec.~\ref{sec:single_defect}). 
The work done on the interface is $ W_{\rm dra} = -\rho g_e h_{\rm dra}^{+} \int\limits_{-\infty}^\infty d x\, \eta_{\rm dra}(x)$,
where the deformation of the interface before the jump, $\eta_{\rm dra}(x)=h_{\rm dra}^{-}-h_{\rm dra}^{+}$. The change of internal energy is
 \begin{align}
\Delta U_{\rm dra} &= U\{h_{\rm dra}^{+}\} - U\{h_{\rm dra}^{-}\} 
\nonumber\\
& = -\int\limits_{-\infty}^\infty 
dx \left[\frac{\gamma}{2} \left(\frac{d
  \eta_{\rm dra}(x)}{d x}\right)^2 +\frac{\rho g_e}{2} (\eta_{\rm dra}^2+2 h_{\rm dra}^{+}\eta_{\rm dra}) -\delta p_c^* \int_{h_{\rm dra}^{+}}^{h_{\rm dra}^{+} + \eta_{\rm dra}} dy\, F(x,y) \right].
\end{align} 
This provides the following dissipated energy:
 \begin{align}
\label{eq:diss_dra}
\Psi_{\rm dra} = \Delta U_{\rm dra}-W_{\rm dra} = -\int_{-\infty}^{\infty}
dx \left[\frac{\gamma}{2}\left(\frac{d
  \eta_{\rm dra}(x)}{d x}\right)^2 + \frac{\rho g_e}{2} \eta_{\rm dra}^2 - \delta
p_c^* \int_{h_{\rm dra}^{+}}^{h_{\rm dra}^{+} + \eta_{\rm dra}} dy\, F(x,y) \right].
\end{align}

\subsection{Energy Dissipation: Mesa defect}

For imbibition in a mesa defect, in the rightmost term in Eq.~\eqref{eq:diss_dif}
\begin{equation}
\int_{y_d}^{y_d + \eta_{\rm imb}} dy\, F(x,y)={\rm Rect}(x/w)\eta_{\rm imb}(x), 
\end{equation}
which, together with the fact that $\eta_{\rm imb}$ equals $\eta$ from Eq.~\eqref{eq:mesa_solution}, allows us to calculate the integral in Eq.~(\ref{eq:diss_dif}), giving
\begin{equation}
\label{eq:psi_imb}
% \dbar 
\Psi_{\rm imb} = -\frac{1}{2}\frac{\delta p_c^{*2} \ell_c}{\rho g_e} \left[\frac{w}{\ell_c} +
\exp(-w/\ell_c) - 1 \right].
%\label{eq:diss_dif_imb}
\end{equation}

Similarly, for drainage we substitute in Eq.~\eqref{eq:diss_dra} $h_{\rm dra}^{+}$ for $h_c$ from Eq.~\eqref{eq:hc}, to obtain 
\begin{equation}
\int_{h_{\rm dra}^{+}}^{h_{\rm dra}^{+} + \eta_{\rm dra}} dy\, F(x,y)={\rm Rect}(x/w)[\eta_{\rm dra}(x)+h_c-y_d].
\end{equation}
Using $\eta$ from Eq.~(\ref{eq:mesa_solution}) for $\eta_{\rm dra}$ in Eq.~\eqref{eq:diss_dra} gives
\begin{equation}
\label{eq:psi_dra}
% \dbar 
\Psi_{\rm dra} = 
%- \frac{1}{2}\frac{\delta p_c^{*2} \ell_s}{\rho g_e} \left[\frac{w}{\ell_s} +
%\exp(-w/\ell_s) - 1 \right] - w \delta p_c (H_c - d). 
-\frac{1}{2}\frac{\delta p_c^{*2} \ell_c}{\rho g_e} \left[1- \left(1 + \frac{w}{\ell_c}\right) \exp(-w/\ell_c)\right].
\end{equation}
The total dissipated energy for a closed hysteresis cycle, 
% $\dbar \Psi_o = \dbar \Psi_{\rm imb} + \dbar \Psi_{\rm dra}$, 
$\Psi_{\rm tot} = \Psi_{\rm imb} +  \Psi_{\rm dra}$, 
is given by 
\begin{equation}
\label{eq:psi_total}
% \dbar 
\Psi_{\rm tot} = -\frac{w\delta p_c^{*2}}{2 \rho g_e}
\left[1 - \exp(-w/\ell_c)\right].
\end{equation}

In the narrow defect limit, $w\ll\ell_c$, we get
\begin{equation}
\label{eq:psi_imb_mesa_narrow}
\Psi_{\rm imb}\approx\Psi_{\rm dra}\approx -\frac{1}{4}\frac{\delta p_c^{*2} w^2}{\rho g_e \ell_c}.
\end{equation}
This dependence on system parameters is expected when analyzing the terms in Eqs.~\eqref{eq:psi_imb} and \eqref{eq:psi_dra} separately, as they are all of the same order of magnitude. 
In particular, the dependence on the defect width $w$ is quadratic, as $\eta_m \sim w$ 
[fixing all other parameters in Eq.~\eqref{eq:psi_imb_mesa_narrow}] and the {width of the} region where $\eta \approx \eta_m$ is $w$-independent.

For the wide-defect limit, $w\gg\ell_c$, dissipation in imbibition reduces to
\begin{equation}
\label{eq:psi_dra_mesa_narrow}
\Psi_{\rm imb}\approx -\frac{1}{2}\frac{\delta p_c^{*2} w}{\rho g_e}.
\end{equation}
Here, the dependence on $w$ is linear and matches those of the second and third terms in Eq.~\eqref{eq:psi_imb}; this is because $\eta_{\rm imb}$ {in the wide-defect limit is (i) approximately constant in the region of width $\approx w$, and (ii) that constant is $w$-independent} [fixing all other parameters in Eq.~\eqref{eq:psi_dra_mesa_narrow}]. 
The first term in Eq.~\eqref{eq:psi_imb} is negligible (as {the integrand} peaks near the edges of the defect in regions of width $\sim \ell_c\ll w$).

For drainage {in the wide-defect limit}, the dissipation is
\begin{equation}
\Psi_{\rm dra}\approx -\frac{1}{2}\frac{\delta p_c^{*2} \ell_c}{\rho g_e}.
\end{equation}
This is width-independent, and considerably smaller than in imbibition ($\Psi_{\rm imb}$), suggesting that the second and third terms {approximately} cancel out. 
Thus, in the wide defect limit, dissipation can be thought of as coming from the edges of the defect, rather than its whole width.
This can also be seen 
%can be explained \hl{maybe confirmed? it's not really an explanation} 
by examining the out-of-equilibrium pressure ($p_e$) during a jump, which is used in Eqs.~\eqref{diff_dissip_irreversible}--\eqref{total_dissip_irreversible} to compute the dissipation as a force-displacement product; 
this is illustrated in Video S1 in {\color{blue}
%\href{https://www.dropbox.com/scl/fo/rkfy7j5xr8ofqr76m7avi/h?rlkey=cmgixv6dhfh01s9b2p85qrn5e&dl=0}
{Supplementary Information (SI)}}.
%show the evolution of the (out-of-equilibrium) pressure $p_e$ during the jump in a mesa defect in imbibition and drainage. 
The imbalance at the edges in drainage implies that it is the source for a appreciable portion  
of the dissipation. 
We note that as the example in Video S1 is for a {moderately-wide defect} (not the wide-defect limit), a substantial part of the dissipation is associated with the motion of the central parts of the interface.

\subsection{Energy Dissipation: Regular defect}

For regular defects, the last term in the dissipation calculations, Eqs.~\eqref{eq:diss_dif} and \eqref{eq:diss_dra}, becomes more complicated, because the {defect consists of two parts}. With $F(x,y)$ given by Eq.~\eqref{pcxz}, this term is
\begin{equation}
\label{eq:Fregimb}
\int_{y_d}^{y_d+\eta_{\rm imb}}dy\, F(x,y)={\rm Rect}(x/w)\begin{cases}
\eta_{\rm imb}^2/(2\ell_s), & \eta_{\rm imb}<\ell_s,\\
\eta_{\rm imb}-\ell_s/2, & \eta_{\rm imb}>\ell_s,
\end{cases}
\end{equation}
for imbibition, and 
\begin{equation}
\label{eq:Fregdra}
\int_{h_{\rm dra}^{+}}^{h_{\rm dra}^{+}+\eta_{\rm dra}}dy\, F(x,y)={\rm Rect}(x/w)\begin{cases}
(\eta_{\rm dra}+h_{\rm dra}^{+}-y_d)^2/(2\ell_s), & \eta_{\rm dra}<\ell_s+y_d-h_{\rm dra}^{+},\\
\eta_{\rm dra}+h_{\rm dra}^{+}-y_d-\ell_s/2, & \eta_{\rm dra}>\ell_s+y_d-h_{\rm dra}^{+}
\end{cases}
\end{equation}
for drainage.

\begin{figure}
\includegraphics[width=.5\textwidth]{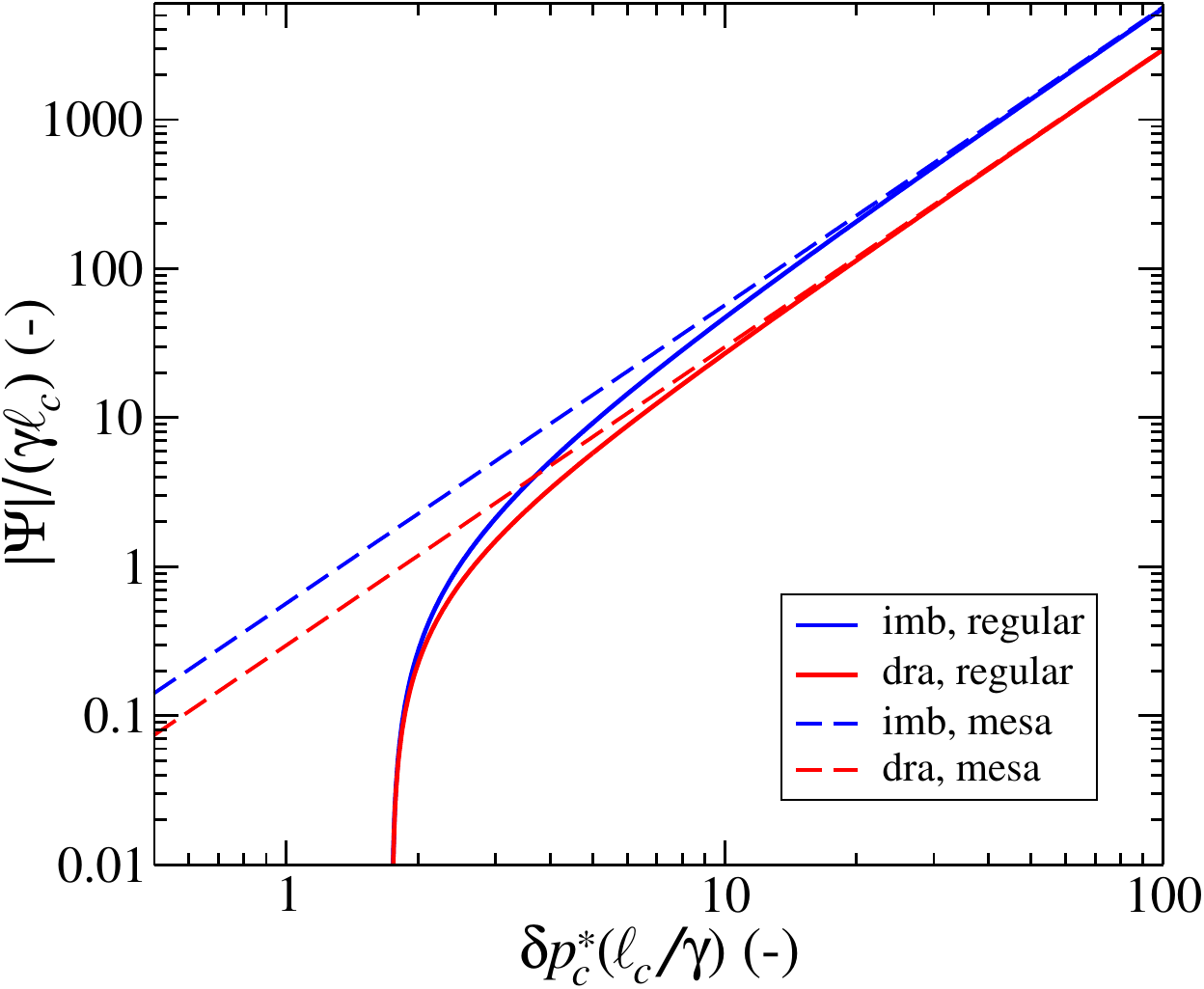}
\caption{Dimensionless dissipated energy against dimensionless capillary {pressure} perturbation for mesa
and regular defects with $w/\ell_c = 2$ ($\ell_s/\ell_c=1$ for regular defects). For very strong defects (large $\delta p_c^*$), the results for mesa and regular defects converge. %, however, there is a threshold for regular defects, but not for mesa defects. 
The difference between imbibition and drainage, which is significant for mesa defects as well as for very strong regular defects, vanishes as $\delta p_c$ approaches the threshold between weak and strong. 
}
\label{fig:dissip}
\end{figure}

Calculating dissipation analytically (or even finding the expression for $\eta$) for regular defects of an arbitrary width and slope is considerably more difficult than for mesa defects. This is because the solution for the interface can be in both the ramp and the plateau parts of the defect, requiring matching between all the different parts of the interface. 
Numerical results for $\eta_{\rm imb}$ and $\eta_{\rm dra}$ can be obtained as described in Appendix~\ref{appendix_profiles}, followed by numerical integration to obtain $\Psi_{\rm imb}$ and $\Psi_{\rm dra}$. 
Nonetheless, analytical results can be obtained for specific cases. 
First, as there is no dissipation for weak defects, the dissipated energy for both imbibition and drainage vanishes as the defect strength approaches the limit $\Pi/\Pi_c^{(1)}\to 1+$. On the other hand, for $\Pi/\Pi_c^{(1)}\gg 1$, the interface solution {crosses the defect} entirely within the plateau region, and therefore $\eta_{\rm imb}$ and $\eta_{\rm dra}$ equal those for a mesa defect {with the same} $w$ and $\delta p_c^*$.
Therefore, the first two terms in Eqs.~\eqref{eq:diss_dif} and \eqref{eq:diss_dra} are {identical for mesa and regular defects}. 
%As for the third term, 
If the interface jumps far into the plateau region, the third term  resembles that 
in the mesa case (e.g., for imbibition $\eta_{\rm imb}-\ell_s/2\approx\eta_{\rm imb}$); if it is not {(as is the case for narrow defects in drainage, see Fig.~\ref{Fig:profiles}a)}, the term is negligible. 
Therefore, the dissipation for $\Pi/\Pi_c^{(1)}\gg 1$ is approximately {equal} to that in a mesa defect (with the same $w$ and $\delta p_c^*$),
\begin{equation}
\Psi_{\rm imb}\approx -\frac{1}{2}\frac{\delta p_c^{*2} \ell_c}{\rho g_e} \psi_{\rm imb} = -\frac{\ell_s^2 \ell_c^3\Pi^2}{2\gamma}
%\left[\frac{w}{\ell_c}+\exp(-w/\ell_c)-1\right]
\psi_{\rm imb}
\label{imbmesa}
\end{equation}
and
\begin{equation}
\Psi_{\rm dra}\approx -\frac{\ell_s^2 \ell_c^3\Pi^2}{2\gamma}
%\left[1-\left(1+\frac{w}{\ell_c}\right)\exp(-w/\ell_c)\right].
\psi_{\rm dra}.
\label{dramesa}
\end{equation}
Here, 
\begin{equation}
\psi_{\rm imb} = \frac{w}{\ell_c}+\exp(-w/\ell_c)-1
\label{imbmesa_brackets}
\end{equation}
and
\begin{equation}
\psi_{\rm dra} = 1-\left(1+\frac{w}{\ell_c}\right)\exp(-w/\ell_c).
\label{dramesa_brackets}
\end{equation}
Both the approach to zero dissipation as $\Pi\to\Pi_c^{(1)}$, and the approach to the mesa results for $\Pi/\Pi_c^{(1)}\gg 1$, are confirmed by Fig.~\ref{fig:dissip}, where numerical computations for a regular defect of an intermediate width ($w/\ell_c=2$) are compared to analytical mesa results.

Similarly, it is easy to calculate the dissipation for a regular defect with an arbitrary slope $\Pi$, in the narrow defect limit ($w\ll\ell_c$). 
The dissipation, {that turns out to be} equal in imbibition and drainage, 
is obtained using Eqs.~\eqref{eq:narrow_imbm}--\eqref{eq:narrow_drap} for the interface shapes [$h_{\rm imb}^{-}$, $h_{\rm imb}^{+}$, $h_{\rm dra}^{-}$ and $h_{\rm dra}^{+}$] in Eqs.~\eqref{eq:diss_dif}, \eqref{eq:diss_dra}, \eqref{eq:Fregimb} and \eqref{eq:Fregdra}, as well as the expression for $\Pi_n^{(1)}$ given by Eq.~\eqref{eq:crit_slope_weak_single}:
\begin{equation}
\Psi_{\rm imb}=\Psi_{\rm dra}=-\frac{w\ell_s^2}{2}\Pi\left(\frac{\Pi}{\Pi_n^{(1)}}-1\right)=-\frac{\ell_s^2 \ell_c^3}{4\gamma}\Pi(\Pi-\Pi_n^{(1)})\left(\frac{w}{\ell_c}\right)^2.
\label{narrow}
\end{equation}
This equals the dissipation for the mesa defect in Eqs.~(\ref{imbmesa})--(\ref{dramesa}) {when both} (i) $w\ll \ell_c$, in which case we can approximate the expressions 
%in square brackets 
in 
Eqs.~(\ref{imbmesa_brackets})--(\ref{dramesa_brackets}) by $(1/2)(w/\ell_c)^2$;
and (ii) $\Pi\gg\Pi_n^{(1)}$, when $\Pi(\Pi-\Pi_n^{(1)})\approx \Pi^2$ in Eq.~(\ref{narrow}).

\begin{figure}
\begin{center}
\includegraphics[width=0.6\linewidth]{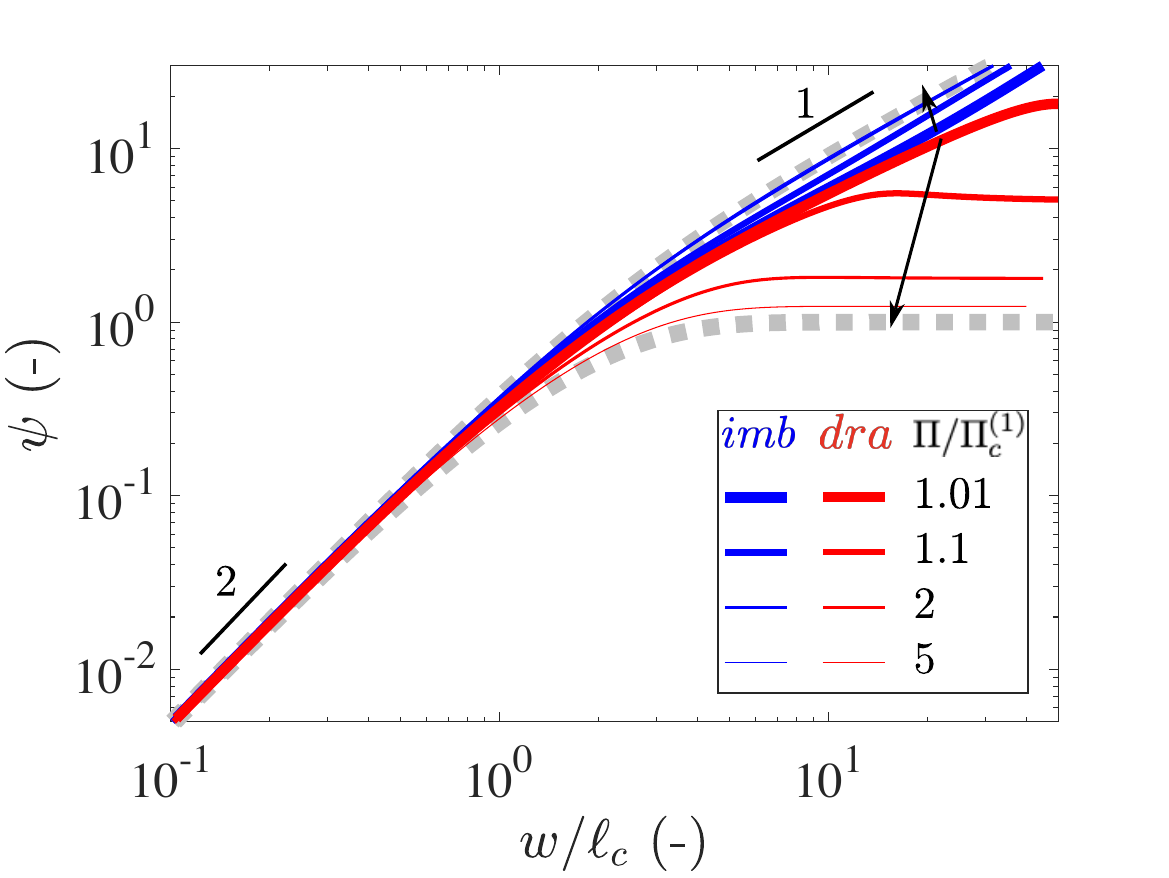}
\end{center}
\caption{
Rescaled dissipated energy $\psi$ [Eq.~(\ref{psi})] in imbibition (blue) and drainage (red) for various values of $\Pi/\Pi_c^{(1)}$ (a single regular defect). 
For narrow defects ($w / \ell_g \ll 1$), the imbibition and drainage curves collapse (independently of $\Pi/\Pi_c^{(1)}$), with $\psi \propto w^2$. For arbitrary widths, for small slopes i.e. $\Pi/\Pi_c^{(1)}\to 1$, imbibition and drainage approach the same limit, whereas as the slope increases (``stronger'' defect; black arrows),  $\Pi/\Pi_c^{(1)}\to\infty$, the dissipation approaches the bounding values of the dissipation in a mesa defect (dashed gray), given by the expressions in Eqs.~(\ref{imbmesa_brackets})--(\ref{dramesa_brackets}). For imbibition at very wide defects, dissipation scales with the width, $\psi \propto w^1$. 
}
\label{psiplot}
\end{figure}

It is useful to consider the following dimensionless quantity,
\begin{equation}
\psi=-\frac{\Psi}{\ell_s^2 \ell_c^3 \Pi(\Pi-\Pi_c^{(1)})/(2\gamma)}.\label{psi}
\end{equation}
For arbitrary defect widths and slopes, Eqs.~\eqref{eq:pc1gen}--\eqref{eq:etadragen} predict that $\psi$ depends only on the dimensionless parameters $w/\ell_c$ and $\Pi/\Pi_c^{(1)}$. 
According to Eq.~\eqref{narrow}, for narrow defects it is expected to depend only on $w/\ell_c$ (and be identical for imbibition and drainage), while for an arbitrary width $w$ it should approach the expressions in
%in the square brackets in 
Eqs.~(\ref{imbmesa_brackets})--(\ref{dramesa_brackets}) as $\Pi/\Pi_c^{(1)}\to\infty$. 
This is demonstrated by plotting the dissipation for various defect widths $w$ and $\Pi/\Pi_c^{(1)}$ (Fig.~\ref{psiplot}).
%, where each curve corresponds to a fixed ratio $\Pi/\Pi_c^{(1)}$ and $w$ is varied, we see that this is indeed the case. 
Figure~\ref{psiplot} also shows that as $\Pi/\Pi_c^{(1)}\to 1$, $\psi$ approaches a finite value, and therefore the dissipation is linear in $\Pi-\Pi_c^{(1)}$ just above the threshold; this was shown in Eq.~(\ref{narrow}) for narrow defects, here confirmed for an arbitrary width. 
This linear dependence is a consequence of the finite interfacial jump; a faster approach to zero is expected when this is not so. 
Moreover, the limits for $\psi$ as $\Pi/\Pi_c^{(1)}\to 1$ are the same for imbibition and drainage, see also Fig.~\ref{fig:dissip}.
%, where the curves merge near the threshold). 
Therefore, the dissipation for imbibition and drainage is similar for narrow defects (regardless of strength $\Pi$) and for arbitrary widths when $\Pi-\Pi_c^{(1)}\ll \Pi_c^{(1)}$.

The dependence of dissipation on the defect width for a regular defect is similar to that for a mesa defect for fixed $\Pi\gg\Pi_c^{(1)}$. 
This can be explained by considerations similar to those we used for the mesa defects. For narrow defects, all the terms in Eqs.~\eqref{eq:diss_dif} and \eqref{eq:diss_dra} are of the same order of magnitude and $\propto w^2$. 
For wide defects, the second and third terms are $\propto w$, giving rise to that dependence of $\Psi$ on $w$ in imbibition, but they cancel out for drainage resulting in $w$-independent dissipation. However, in contrast to mesa defects, for regular defects dissipation also vanishes as $\Pi\to\Pi_c^{(1)}$, i.e. {the terms that are of the same order of magnitude must} cancel out.
This is easy to confirm for imbibition in wide defects, where the first term is negligible and the last two terms in the integrand are constant within the defect not too close to its edges. 
The sum of these terms is
\begin{equation}
\frac{\rho g_e}{2}\eta_{\rm imb}^2-\delta p_c^* \int_{y_d}^{y_d + \eta_{\rm imb}} dy\, F(x,y)=\begin{cases}
(1/2)(\rho g_e-\Pi)\eta_{\rm imb}^2, & \eta_{\rm imb}<\ell_s,\\
(1/2)(\rho g_e-\Pi)\eta_{\rm imb}^2+(\Pi/2)(\eta_{\rm imb}-\ell_s)^2, & \eta_{\rm imb}>\ell_s,
\end{cases}
\end{equation}
where we have used Eq.~\eqref{eq:Fregimb}. 
This has the incorrect (positive) sign for any $\eta_{\rm imb}>0$ when $\Pi<\Pi_w^{(1)}=\rho g_e$, thus, there are no nonzero solutions for $\eta_{\rm imb}$, i.e., {no stable deformed configurations, and the defect is weak}. For a strong defect
[$\Pi>\Pi_w^{(1)}$] and $\eta_{\rm imb}=\ell_s(\Pi/\Pi_w^{(1)})$ [see Eqs.~\eqref{eq:wide_imbm}--\eqref{eq:wide_imbp}], this sum is negative and approaches zero when $\Pi\to\Pi_w^{(1)}$, as expected. 
We can also verify that the last two terms in the expression for drainage cancel out for \textit{any} $\Pi>\Pi_w^{(1)}$. For $\eta>\ell_s+y_d-h_{\rm dra}^{+}$ the sum of these terms becomes 
\begin{equation}
\frac{\rho g_e}{2}\eta_{\rm imb}^2-\delta
p_c^* \int_{h_{\rm dra}^{+}}^{h_{\rm dra}^{+} + \eta_{\rm dra}} dy\, F(x,y)=\frac{\Pi_w^{(1)}}{2}\eta_{\rm dra}^2-\Pi\ell_s(\eta_{\rm dra}+h_{\rm dra}^{+}-y_d-\ell_s/2),
\end{equation}
where we used Eq.~\eqref{eq:Fregdra}. 
Using Eq.~\eqref{eq:wide_dram} for $h_{\rm dra}^{-}=\eta_{\rm dra}+h_{\rm dra}^{+}$ and Eq.~\eqref{eq:wide_drap} for $h_{\rm dra}^{+}$, we see that the sum vanishes for all $\Pi$ values.

%%%%%%%%%%%%%%%%%%%%%%%%%%%%%%%%%%%%%%%
%%%%%%%%%%%%%%%%%%%%%%%%%%%%%%%%%%%%%%%
%%%%%%%%%%%%%%%%%%%%%%%%%%%%%%%%%%%%%%%
\section{cooperative origin of hysteresis and dissipation}
\label{sec:cooperative_dissipation} 
%%%%%%%%%%%%%%%%%%%%%%%%%%%%%%%%%%%%%%%
%%%%%%%%%%%%%%%%%%%%%%%%%%%%%%%%%%%%%%%
%%%%%%%%%%%%%%%%%%%%%%%%%%%%%%%%%%%%%%%

A single mesa defect is strong, with interface displacement (e.g. PS trajectory) which is hysteretic and irreversible (dissipative). 
A disordered medium composed of multiple mesa defects, is also hysteretic and dissipative \cite{Holtzman2023}. 
From this, one might naively conclude that the overall hysteretic response is simply the sum of the response of basic hysteretic entities (here, ``defects''). 
This conforms with the conventional thinking behind compartment models such as Leverett and Preisach (where basic hysteretic entities are called ``hysterons'') \cite{Bertotti_V1}.
Following the same logic, one would expect that a medium composed of multiple weak defects---each of which is non-hysteretic, reversible and non-dissipative---will also be non-hysteretic and reversible.
To examine this, % we explore here the interaction between a pair of weak regular defects and its effect on the passing interfaces.
% \cite{Joanny1984}
%This is not the case when dealing with weak defects, where single defects do not constitute elementary units of dissipation and hysteresis. So that, we explore here the interaction between weak defects \cite{Joanny1984}, and their effect on passing interfaces.
we consider a simple system composed of two identical regular, weak defects of width $w$ at a distance $d$ apart (measured between their centers, with $d\ge w$), positioned at $y=y_d$ (Fig~\ref{Fig1_sketch_setup}). 
Similarly to the single defects considered in Sections.~\ref{sec:config_single} and \ref{sec:energy_balance}, each defect has a capillary pressure profile which is linearly increasing in $y$ over a distance $\ell_s$, after which it remains constant (plateau), with the total defect length $\ell$. 
For each system with a given interdefect distance $d$, we compute the energy dissipated in imbibition and drainage, analytically and numerically, as described below.

%%%%%%%%%%%%%%%%%%%%%%%%%%%%%%%%%%%%%%%
%%%%%%%%%%%%%%%%%%%%%%%%%%%%%%%%%%%%%%%
\subsection{Interactions between a pair of weak defects: Analytical evaluation}
%of the interactions between a pair of weak defects}
%%%%%%%%%%%%%%%%%%%%%%%%%%%%%%%%%%%%%%%
%%%%%%%%%%%%%%%%%%%%%%%%%%%%%%%%%%%%%%%

The local pressure balance in Eq.~\eqref{eq:pressure_balance_2} is general, and thus it can be used for any number of defects, regular or irregular. For the pair of weak defects considered here, we write the capillary pressure profile as
\begin{equation}
\label{eq:F2}
F(x,y)=a_2(x)c(y),
\end{equation}
where $c(y)$ has {the same} form as in the single regular defect case [Eq.~\eqref{eq:cy}], {with} the term $a_1(x)$ [Eq.~\eqref{eq:ax}] replaced here with $a_2(x)$ consisting of two rectangular functions,
\begin{equation}
\label{eq:a2}
a_2(x)={\rm Rect}\left(\frac{x+d/2}{w}\right)+{\rm Rect}\left(\frac{x-d/2}{w}\right)=w \left[\delta_w(x + d/2) + \delta_w(x - d/2)\right],
\end{equation}
where we use $\delta_w(x)$ introduced in Section~\ref{sssec:system_narrow}. Note that for $d = w$, the two-defect system reduces to a single defect of width $2 w$. 

Consider the case when the defects are narrow ($w/\ell_c\ll 1$), so that we can replace $\delta_w(x)\to\delta(x)$. Using the resulting approximation of $a_2(x)$ in~\eqref{IE} (substituting $a_2$ for $a_1$), under the assumption that the interface solution {crosses the defects} within the ramps, the deformation is
\begin{equation}
\eta(x) = \Pi w [G_0(x+{d/2}) + G_0(x - {d/2}) ]
[(h_0 - y_d) + \eta_1], 
\label{eq:eta_2def}
\end{equation}
where we set $\eta_1 = \eta(-{d/2}) = \eta({d/2})$ by using the
symmetry of the domain. 
Imposing $x = {d/2}$ in \eqref{eq:eta_2def}, we obtain 
\begin{equation}
\eta_1 = \frac{\Pi w \beta_1 (h_0 - y_d) }{1 - \Pi w \beta_1} \quad \mbox{with}\quad \beta_1 =  G_0(d) + G_0(0)
\end{equation}
%$
and therefore 
\begin{equation}
\eta(x) = \left[G_0(x+{d/2}) + G_0(x - {d/2}) \right]
\frac{\Pi w (h_0 - y_d)}{1- \Pi w \beta_1}.  
\end{equation}
This is consistent with our assumption that the interface solution {crosses the defects} within the ramps for sufficiently small positive $h_0-y_d$, when the denominator is positive, i.e.,
\begin{equation}
\label{ac}
\Pi<\Pi_n^{(2)} = \frac{1}{w \beta_1} = \frac{1}{w \beta_0 [\exp(-{d}/\ell_c)
  + 1]} = \frac{\Pi_n^{(1)}}{\exp(-d/\ell_c)
  + 1} \leq \Pi_n^{(1)}.
\end{equation}
As for the single defect, this is the criterion for weakness of the pair of defects, and $\Pi_n^{(2)}$ is the (narrow-defect approximation of) the critical slope for the transition from weak to strong. 
This method can be extended to an arbitrary number of defects which can also have different defect strengths (disordered media). 
Note that this critical slope for the two-defect system is always smaller than the one for the single defect, $\Pi_n^{(1)}$. This holds for defects of arbitrary width; for derivation of the critical slope $\Pi_c^{(2)}$ {for this case} see Appendix~\ref{appendix_critical_slope}.
For slopes within $\Pi_c^{(2)} < \Pi < \Pi_c^{(1)}$, each defect is
non-dissipative (weak) when isolated (single defect only), whereas a system of two such defects is dissipative and hysteretic (strong).

%\paragraph{Critical distance for fixed slope}
Within the narrow-defect approximation, 
let us consider now a situation where the slope of the individual defects is $\Pi < \Pi_n^{(1)}$, that is, each defect by itself is weak. 
The two-defect system is strong if $\Pi>\Pi_n^{(2)}$, implying that 
\begin{align}
\left[\exp(-d /\ell_c) + 1\right] > \frac{\Pi_n^{(1)}}{\Pi}.
\end{align}
This inequality indicates that the two-defect system becomes strong if the separation distance $d$ between the defect centers is $w \leq d < d_c$,
where
\begin{equation}
d_c =  \ell_c
\ln\left(\frac{\Pi}{\Pi_n^{(1)} - \Pi} \right).  
\label{Eq_d_crit}
\end{equation}
This is meaningful only for $d_c > w\approx 0$, that is, for
$\displaystyle\frac{\Pi_n^{(1)}}{2} \leq \Pi < \Pi_n^{(1)}$. 
Thus, {there is a factor of two} between the largest and smallest slopes where a single defect is weak but a pair can be strong. 
The factor of two reduces for wider defects; in the limit $w/\ell_c\gg 1$, it approaches unity, as the critical slope approaches $\rho g_e=\gamma/\ell_c^2$, independent of $d$.

Considering the case when the interface solution {crosses the defects} within the plateaus, Eq.~\eqref{IE0} (with $a_1$ replaced by $a_2$) gives for narrow defects
\begin{equation}
\label{eq:eta_pair_narrow}
\eta(x)=\Pi w\ell_s[G_0(x+d/2)+G_0(x-d/2)].
\end{equation}
At the defects, this gives
\begin{equation}
\label{eq:eta1}
\eta(\pm d/2)=\frac{\Pi w\ell_s\ell_c}{2\gamma}[1+\exp(-d/\ell_c)]=\frac{\Pi}{\Pi_n^{(2)}}\ell_s,
\end{equation}
a result analogous to the single-defect case [see Eq.~\ref{eq:narrow_plateau}]. 
For dissipation calculations, since the width of the defect cannot be neglected, we write 
\begin{equation}
\eta(x)=\frac{\Pi}{\Pi_n^{(2)}}\ell_s
\begin{cases}
\exp\{-[|x|-(d+w)/2]/\ell_c\}, & |x|>(d+w)/2,\\
\cosh[x/\ell_c]/\cosh[(d-w)/(2\ell_c)], & |x|<(d-w)/2,\\
1, & (d-w)/2<|x|<(d+w)/2,
\end{cases}
\end{equation}
{where we have assumed that $\eta(x)$ is constant within the defects, and equal to Eq.~\eqref{eq:eta1}, and the first two lines are, essentially, Eq.~\eqref{eq:eta_pair_narrow}, with $d$ replaced by $d+w$ in the first line and $d-w$ in the second (a negligible change) to make the result continuous.} 
If the pair is strong ($\Pi/\Pi_n^{(2)}>1$), during imbibition the interface experiences a jump when it first touches the defects, just as in the single-defect case; thus, the interface configurations before and after the jump are
\begin{equation}
\label{eq:himbmpair}
h_{\rm imb}^{-}=y_d
\end{equation}
and
\begin{equation}
\label{eq:himbppair}
h_{\rm imb}^{+}=y_d+\frac{\Pi}{\Pi_n^{(2)}}\ell_s
\begin{cases}
\exp\{-[|x|-(d+w)/2]/\ell_c\}, & |x|>(d+w)/2,\\
\cosh[x/\ell_c]/\cosh[(d-w)/(2\ell_c)], & |x|<(d-w)/2,\\
1, & (d-w)/2<|x|<(d+w)/2.
\end{cases}
\end{equation}
Likewise, during drainage the interface behaves similarly to the single-defect case, where the jump occurs when it reaches the boundary between the plateau and the ramp, with
\begin{equation}
\label{eq:hdrampair}
h_{\rm dra}^{-}=\begin{cases}
y_d-\ell_s\left\{\frac{\Pi}{\Pi_n^{(2)}}(1-\exp\{-[|x|- \frac{d+w}{2} ]/\ell_c\})-1\right\}, & |x|>(d+w)/2,\\
y_d-\ell_s\left\{\frac{\Pi}{\Pi_n^{(2)}}(1-\cosh[\frac{x}{\ell_c}]/\cosh[\frac{d-w}{2\ell_c}])-1\right\}, & |x|<(d-w)/2,\\
y_d+\ell_s, & (d-w)/2<|x|<(d+w)/2
\end{cases}
\end{equation}
and
\begin{equation}
\label{eq:hdrappair}
h_{\rm dra}^{+}=y_d-\ell_s(\Pi/\Pi_n^{(2)}-1).
\end{equation}
The corresponding calculations of interface shapes for defects of an arbitrary width {can be done numerically using the method in} Appendix~\ref{appendix_profiles}.
%, interface shapes before and after jumps can be calculated as described in Appendix~\ref{appendix_profiles}.

The dissipated energy in imbibition and drainage can be evaluated using eqs.~\eqref{eq:diss_dif} and \eqref{eq:diss_dra}, with the values of $h_{\rm imb}^{\pm}$ and $h_{\rm dra}^{\pm}$ computed analytically or numerically, 
%We use eqs.~\eqref{eq:diss_dif} and \eqref{eq:diss_dra} remain valid, 
and the expressions for the integrals of $F(x,y)$ from Eqs.~\eqref{eq:Fregimb}--\eqref{eq:Fregdra}, where ${\rm Rect}(x/w)$ are replaced by ${\rm Rect}([x+d/2]/w)+{\rm Rect}([x-d/2]/w)$. 
For the narrow defect approximation, using Eqs.~\eqref{eq:himbmpair}--\eqref{eq:hdrappair}, as well as Eq.~\eqref{ac} for $\Pi_n^{(2)}$, the result reads
\begin{equation}
\label{eq:Psi_pair}
\Psi_{\rm imb}=\Psi_{\rm dra}=-w\ell_s^2\Pi\left[\frac{\Pi}{\Pi_n^{(2)}}-1\right].
\end{equation}
This expression resembles its counterpart for the single-defect case, Eq.~\eqref{narrow}, except for the critical threshold at which dissipation approaches zero [$\Pi_n^{(2)}$ instead of $\Pi_n^{(1)}$] and the factor of two; this is intuitive, as for two defects far apart the dissipation is additive, and the threshold remains {the same}. 
{Equation~\eqref{eq:Psi_pair} is also} consistent with the fact that two touching defects ($d=w$) are equivalent to a single defect of twice the width.
Finally, we find that the accuracy of the narrow-defect theory can be improved upon replacing in Eq.~\eqref{eq:Psi_pair} the approximate threshold, $\Pi_n^{(2)}$, with the exact result, $\Pi_c^{(2)}$ [Eqs.~\eqref{eq:pic2vszeta}--\eqref{eq:tanzeta}], providing
\begin{equation}
\Psi_{\rm imb}=\Psi_{\rm dra}=-wl^2 \Pi \left[\frac{\Pi}{\Pi_c^{(2)}}-1\right].
\label{narrowpair}
\end{equation}

\subsection{Interactions between a pair of weak defects: Numerical verification}

To examine these intriguing theoretical predictions, we use numerical computations and simulations varying the distance $d$ for a fixed defect shape (slope $\Pi$, {ramp length $\ell_s$, width $w$}), computing the energy dissipated during the imbibition and drainage trajectories
(see Appendix~\ref{appendix_numerical} for parameter values).
We calculate the dissipated energy using (i) the numerical computations described in Appendix~\ref{appendix_profiles}; (ii) direct numerical simulations of the interface evolution and the corresponding energy dissipation \cite{Holtzman2023} 
    %\hl{RH: need to add appendix with similar context to SI of our GRL 2023?} 
(referred to as simulations, to distinguish from the numerical computations of Appendix~\ref{appendix_profiles}). 
%For details of these simulations see \cite{Holtzman2023}. 
While the numerical simulations are more computational costly than the computations in Appendix~\ref{appendix_profiles}, the simulations can be used for any arbitrary capillary pressure field $p_c(x,y)$ (e.g. disorder with prescribed defect strength distributions in \cite{Holtzman2023}).

\begin{figure}
%\centering
\includegraphics[width=0.55\linewidth]{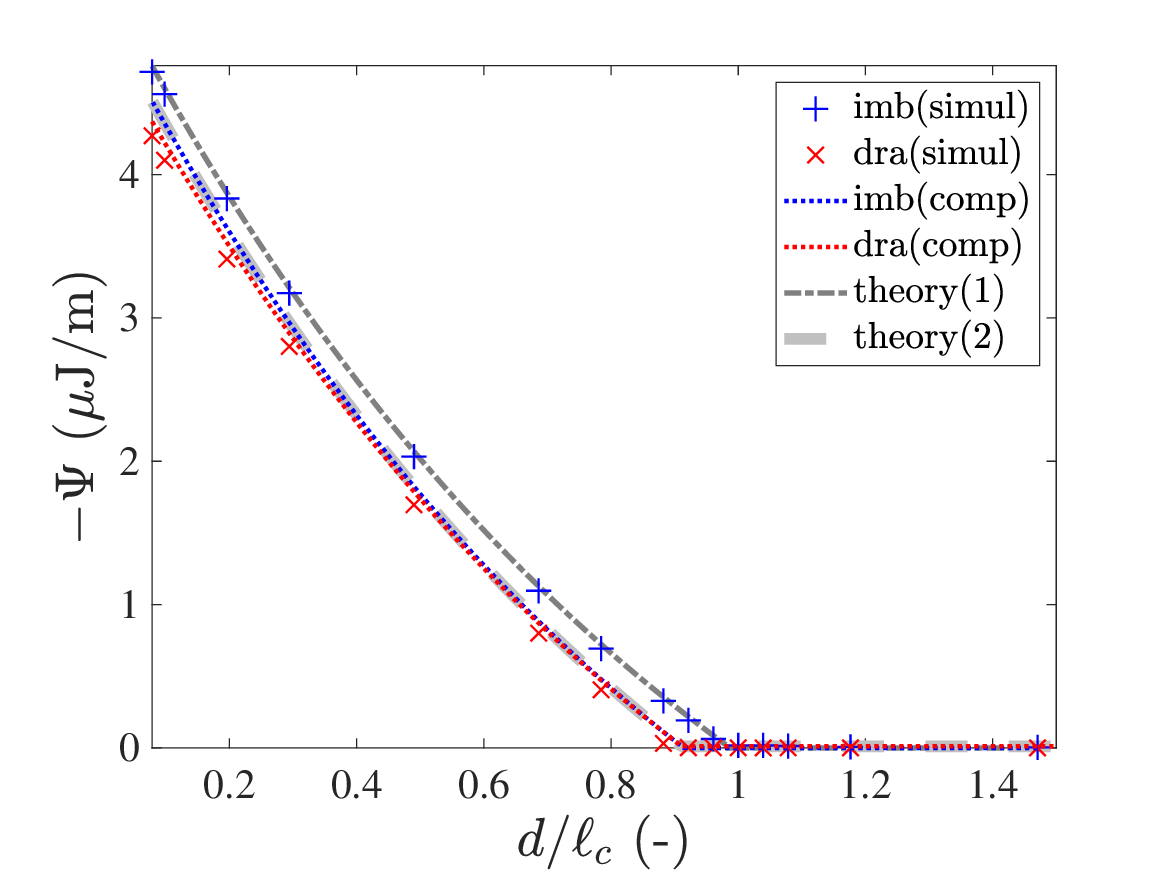}
\caption{
Emergence of energy dissipation (``\emph{strong}'' behavior) in a pair of \emph{weak} defects. As long as the defects are sufficiently far apart ($d > d_c$), the system is reversible and there is no dissipation. 
Once $d<d_c$, the displacement becomes irreversible, and dissipation and hysteresis emerge.
We plot the dissipation calculated from simulations (\cite{Holtzman2023}) and numerical computation (Appendix \ref{appendix_profiles}; dotted lines), where blue and red refer to imbibition and drainage, respectively. 
Also plotted are the analytical solutions from two variants of the narrow defect theory:
%, varying by accuracy due to the way the critical slope is obtained: 
(1) Eq.~\eqref{eq:Psi_pair} (approximate 
%with the approximation in Eq.~\eqref{ac} for 
$\Pi_n^{(2)}$ [dash-dot]; and (2) the more accurate Eq.~\eqref{narrowpair} with an exact value of $\Pi_c^{(2)}$ 
%obtained from Eqs.~\eqref{eq:pic2vszeta}--\eqref{eq:tanzeta} 
[thick dashed]. 
%Vertical lines show the corresponding critical distance $d_c$.
}
\label{Fig:weak_pair}
 \end{figure}

Our numerical evaluations show that for pairs of weak defects sufficiently far apart, $d>d_c$ [where $d_c$ is approximated by Eq.~\eqref{Eq_d_crit}], there is no dissipation and hysteresis (Video S2 in {\color{blue}
%\href{https://www.dropbox.com/scl/fo/rkfy7j5xr8ofqr76m7avi/h?rlkey=cmgixv6dhfh01s9b2p85qrn5e&dl=0}
SI)}, whereas for $d<d_c$, dissipation emerges (Fig.~\ref{Fig:weak_pair}). 
% \hl{Consider adding another plot in the figure to support this, for instance a PS curve or large d and small d.}
This dissipation arises from abrupt jumps of the interface along the defect slope in both imbibition and drainage, such that the PS response becomes hysteretic; the closer the defects are, the stronger the dissipation (and the width of the hysteresis cycle, e.g. see Videos S3--S4 in {\color{blue}
%\href{https://www.dropbox.com/scl/fo/rkfy7j5xr8ofqr76m7avi/h?rlkey=cmgixv6dhfh01s9b2p85qrn5e&dl=0}
{SI}}). 
% [RH: 2 videos for $d<d_c$]

%
%%%%%%%%%%%%%%%%%%%%%%%%%%%%%%%%%%%%%
%%%%%%%%%%%%%%%%%%%%%%%%%%%%%%%%%%%%%

As expected for narrow defects, 
[Eq.~\eqref{eq:Psi_pair}], 
the numerical computations give nearly identical dissipated amounts for imbibition and drainage.
The simulations provide similar results, though slightly exaggerate the difference between imbibition and drainage, due to numerical discretization errors (can be reduced by refinement, at the expense of higher computational cost). 
%These errors can be reduced by improving the resolution of the discretization, at a higher computational cost.
%and seem to predict slightly different thresholds for imbibition and drainage, but this is a numerical discretization artifact; there is otherwise very good agreement between the two approaches. 
The distance at which dissipation vanishes in the numerical computations is identical to the exact result for the critical distance from Eqs.~\eqref{eq:pic2vszeta}--\eqref{eq:tanzeta} 
%(vertical dashed line in 
(``theory(2)'' in legend of Fig.~\ref{Fig:weak_pair}).
The narrow defect theory 
[dissipation from Eq.~\eqref{eq:Psi_pair} and critical distance from Eq.~\eqref{Eq_d_crit}; ``theory(1)'' in legend of Fig.~\ref{Fig:weak_pair}] provides a reasonable approximation. 
For wider defects, the deviation between the dissipation computed in imbibition and drainage, 
%computed with Eq.~\eqref{eq:Psi_pair}, 
and between these and the dissipation evaluated with Eq.~\eqref{narrowpair}, increases.
We note that the close agreement between the simulated dissipation for imbibition and Eq.~\eqref{eq:Psi_pair} is coincidental.

%%%%%%%%%%%%%%%%%%%%%%%%%%%%%%%%%
%%%%%%%%%%%%%%%%%%%%%%%%%%%%%%%%%
\subsection{Interactions between a pair of weak defects: Experimental observation}
%%%%%%%%%%%%%%%%%%%%%%%%%%%%%%%%%
%%%%%%%%%%%%%%%%%%%%%%%%%%%%%%%%%

%{\color{brown}
Here, we provide an experimental proof-of-concept showing the emergence of dissipation in a pair of weak defects as they are brought close to each other. 
We use 3-D printing to manufacture a series of systems (imperfect Hele-Shaw cells), with a single weak defect and two pairs of identical defects at two different separation distances. 
As we cannot measure the energy dissipated experimentally, we use the hysteresis cycle as a proxy for reversibility; the larger the area within a closed PS cycle is, the larger the dissipation  \cite{Holtzman2023}.
%We demonstrate experimentally the cooperative origin of dissipation arising from the interaction of weak defects by 3-D printing a series of imperfect Hele-Shaw cells, with both single defect as well as a pair of defects with varying inter-defect gap  (Fig.~\ref{fig:hysteresis}); 
Details of the experiments including the manufacturing, setup and image analysis, are provided in Appendix \ref{appendix_experimental}.

\begin{figure}
    \centering
    \includegraphics[width=\textwidth]{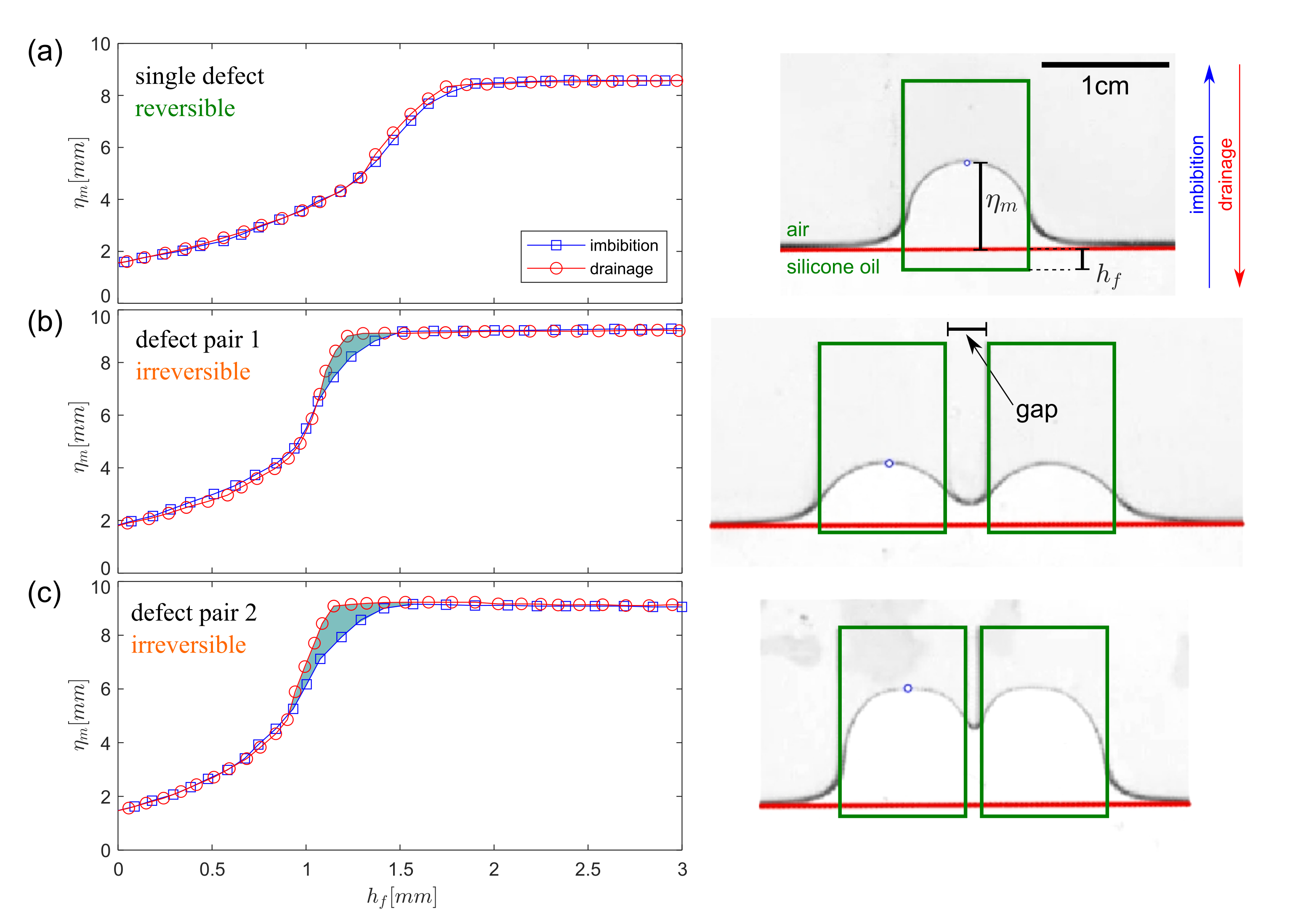}
    \caption{
Imbibition-drainage cycle for a single regular defect (a) and a pair of regular defects of width $w = 10$ mm,
with a separation (measured between their centres) $d$ of (b) 13 mm and (c) 11 mm, respectively. 
For each case, we show the imbibition-drainage trajectories in terms of the interface deformation $\eta_m$ vs. its baseline position $h_f$ (left panels) and an experimental image of the interface (right).
Here $\eta_m$ is the distance between the blue dot and the red line. Defects are highlighted by green rectangles. 
While the interface moves through a single regular defect reversibly, i.e. with no hysteresis and therefore no dissipation (a), a pair of such defects close enough together exhibits hysteresis (b), which increases as their separation distance decreases (c), confirming our theoretical prediction.
}
\label{fig:hysteresis}
\end{figure}

Our experiments validate the findings we obtained theoretically and numerically: while a single regular defect of given geometry (weak) can behave reversibly, showing no hysteresis (and thus no dissipation) (Fig.~\ref{fig:hysteresis}a; see also Video S5 in {\color{blue}
%\href{https://www.dropbox.com/scl/fo/rkfy7j5xr8ofqr76m7avi/h?rlkey=cmgixv6dhfh01s9b2p85qrn5e&dl=0}
SI}), a pair of defects (each of identical geometry to the former) close enough together becomes hysteretic (Fig.~\ref{fig:hysteresis}b; Video S6 in {\color{blue}
%\href{https://www.dropbox.com/scl/fo/rkfy7j5xr8ofqr76m7avi/h?rlkey=cmgixv6dhfh01s9b2p85qrn5e&dl=0}
SI}) due to the spatial interactions between the otherwise reversible entities. Decreasing the pair separation increases the hysteresis (Fig.~\ref{fig:hysteresis}c; Video S7 in {\color{blue}
%\href{https://www.dropbox.com/scl/fo/rkfy7j5xr8ofqr76m7avi/h?rlkey=cmgixv6dhfh01s9b2p85qrn5e&dl=0}
SI}). 
In Fig.~\ref{fig:hysteresis}, we measure the maximum deformation $\eta_m$ along the middle line of the defect, and the baseline position $h_f$ as the vertical distance between the unperturbed interface (far from the defect) and the bottom of the defect, i.e., $h_f = h_0 - y_d$ (see also Fig.~\ref{Fig1_sketch_setup}).

We note that in these experiments, effects that are not considered in our 2-D linear model can be of importance
    \footnote{
    %% Noting also that 
    %In our 2-D model we use an approximation of the total curvature as the sum of in- and out-of-plane components, with the linear approximation for the former, which deviates from the full nonlinear formulation e.g. in \cite{Lavi_PRF_2023} for large deformations. Furthermore, 
    We also note that our 2-D model is strictly valid only when the out-of-plane thickness varies slowly in space, while in the examples considered here the defects contain abrupt changes in thickness.}. 
One such effect is
%large, non-linear interface deformation 
large interface deformation resulting in nonlinear curvature, where in  
    %\footnote
    {our 2-D model we (i) approximate the total curvature as the sum of in- and out-of-plane components, and (ii) use a linear approximation for the former (vs. the full nonlinear formulation, e.g. see \cite{Lavi_PRF_2023})}. 
Further differences between our 2-D model and the experiments arise from 3-D effects related to (i) the curvature of the meniscus between adjacent defects; (ii) the finite width of the meniscus in-plane projection; and (iii) thin liquid film advancing ahead of the experimentally-observed \emph{2-D projection} of the (3-D) meniscus, resulting in an uncertainty in its position, hence $\eta_m > 0$ at $h_f<0$ in Fig.~\ref{fig:hysteresis}; see further elaboration in Appendix \ref{appendix_experimental}.
 These effects preclude a quantitative comparison between our theoretical predictions and the experiments, which are presented here as proof-of-concept that qualitatively supports our theory.

%%%%%%%%%%%%%%%%%%%%%%%%%%%%%%%%%%%%%%%
% \section{Discussion and Conclusions}
\section{Conclusions}
\label{sec:discussion_conclusions}

In this paper, we study energy dissipation during quasi-static fluid-fluid displacements across 
an imperfect Hele-Shaw cell (with ``defects'' i.e. local perturbations in thickness and thus capillary pressure $p_c$).
To explore the fundamental mechanisms for energy dissipation and hysteresis, we consider a simple system comprising of isolated (single) defects of two shapes: ``regular'' with \emph{continuous} (here linear) variations in $p_c$ (in the direction of fluid advancement), and ``mesa'' defects where $p_c$ changes \emph{abruptly}. 
Depending on the slope of $p_c$, the system can be classified as either (i) ``weak'', where the interface passage across the defect (imbibition and drainage) is fully reversible with no dissipation and hysteresis; and (ii) ``strong'', with dissipation and hysteresis.
We derive analytical expressions for the evolution of the fluid-fluid configuration as it deforms when it passes through defects and the corresponding energy dissipation, which are validated numerically. 
The analysis exposes subtle morphological differences between imbibition and drainage, originating from the differences in wet area when a jump occurs.

A novel intriguing finding is that irreversible, hysteretic behavior and the associated energy dissipation can emerge from the interaction of weak defects---objects which are intrinsically (when isolated) non-dissipative and non-hysteretic. 
This is demonstrated for a simple system: a pair of identical weak defects, varying their separation distance.
While far enough apart, the pair of defects behaves as weak. Once the defect distance falls below a threshold---dictated by surface tension and the system parameters (defect width and slope of $p_c$), it becomes strong, producing hysteresis and dissipation. 
We compute this threshold and the energy dissipated analytically, for the approximation of narrow defects (width much smaller than the capillary length), and numerically (for arbitrary width) using two methods: (i) 
%RFIM-like 
numerical simulations of the sequence of equilibrium configurations and (ii) numerical computation of the equilibrium conditions.
A proof-of-concept experiment using 3-D printed cells demonstrates the emergence of irreversible, hysteretic behavior due to the interaction between non-hysteretic, reversible defects, validating our theoretical findings.

The cooperative mechanism exposed here is uniquely highlighted by our model.
In contrast, classical models for hysteresis such as compartment models~\cite{Bertotti_V1} rely on the existence of a basic hysteretic unit (``hysteron''), 
and the overall hysteretic response is simply the sum of the responses of these units. 
In the zero-temperature RFIM isolated spin flips are reversible \cite{SethnaPRL1993}, and thus individual spins play a role equivalent to our weak defects. 
Similarly to the present case, hysteresis and dissipation in the RFIM emerge from the collective response of the system, in the form of spin-flip avalanches triggered by the local spin-spin interactions. 
A crucial difference, however, is that spin-spin interactions in RFIM are present in the whole system, and thus irreversible events can occur anywhere. 
In contrast, the interaction between defects in our system is mediated by the two-phase interface, so that irreversible jumps and dissipation are localized at the interface. 
Interestingly, the emergence of hysteresis as a collective response of a system of individually non-hysteretic agents with continuum responses, interacting at each step through the memory of the predecessors, was applied to explain irreversibility in climate events \cite{Lim2002}.

To the best of our knowledge, our model is the first to describe how hysteresis of pressure-saturation curves during cyclic flows in disordered media emerges from the interactions among defects, in a physical, non-phenomenological manner. 
%\hl{JO: clarify that this is in the present context of PS hysteresis. In the RFIM, for example, the whole hysteretic behaviour is due to the interactions between adjacent spins.} 
Specifically in the context of fluid displacement, both (i) the ``ink bottle effect''---the canonical conceptual model for the pore-scale mechanism underlying pressure-saturation hysteresis \cite{Haines1930}, as well as (ii) the ``Leverett model''---a compartment model predicting macroscopic hysteresis \cite{Bertotti_V1}, do not include the key concept of spatial distance among the basic units (pores or defects). 
An oversimplification of the interactions between basic model units is also inherent to the classical capillary bundle model used in soil physics to predict the pressure-saturation relationship \cite{Hunt_SSSAJ2013}.
We argue that the cooperative mechanism unveiled here is the dominant mechanism responsible
% at the heart 
for hysteresis in multiphase fluid systems, noting that other mechanisms such as contact angle hysteresis in surface wetting  \cite{Robbins1987,Giacomello_PNAS2016}, changes in interfacial connectivity (snap-off) and fluid trapping \cite{Albers2014} also contribute to hysteresis.

In conclusion, we present a detailed, rigorous investigation of the fundamental process of energy dissipation between consecutive metastable configurations in the passage of an interface through topographical defects.
%, by interfacial deformation at 
In the quasi-static limit considered here, viscous dissipation due to finite velocity of the interface displacement is disregarded.
The simplicity of this conceptual model 
% avoiding the complexity of nonlocal interactions among multiple dissipative entities and disorder, 
allows systematic 
%experimental and 
theoretical examination of the origins of energy dissipation 
%and its link to memory 
and hysteresis.
%, helps clarifying key concepts that have so far remained elusive. 
% and controversial. 
%
The insights gained here considering individual defects
% ---the building blocks of a disordered media made of multiple interacting defects---
are of immediate relevance to disordered media containing multiple interacting defects \cite{Holtzman2023}. 
Interesting directions for further studies are the introduction of mechanisms that are not considered in the current model system, to gain understanding of two-phase flow in more complex disordered systems such as porous and fractured materials \cite{maloy2021burst}. 
One is dynamic effects associated with viscous dissipation in rate-driven systems and at high flow rates, connecting insights gained from detailed investigation of single isolated defects \cite{Lavi_PRF_2023}. 
Another is the nonlinear, 3-D effects that were evident even in the simple experiments presented here.
Finally, links between our novel modelling approach and the RFIM and other lattice models open the way to gain fundamental understanding of how cooperative interactions among non-hysteretic, non-dissipative entities could lead to the emergence of hysteresis and dissipation in a wide variety of driven disordered systems \cite{Lindeman2021,Wiese_2022,Shohat_PRL2023}.

%%%%%%%%%%%%%%%%%%%%%%%%%%%%%%%%%%%
%%%%%%%%%%%%%%%%%%%%%%%%%%%%%%%%%%%
%%%%%%%%%%%%%%%%%%%%%%%%%%%%%%%%%%%
%%%%%%%%%%%%%%%%%%%%%%%%%%%%%%%%%%%
%%%%%%%%%%%%%%%%%%%%%%%%%%%%%%%%%%%
%%%%%%%%%%%%%%%%%%%%%%%%%%%%%%%%%%%

%%%%%%%%%%%%%%%%%%%%%%%%%%%%%%%%%%%
%%%%%%%%%%%%%%%%%%%%%%%%%%%%%%%%%%%
\appendix
% Specify following sections are appendices. Use% \appendix* %if there only one appendix.
%%%%%%%%%%%%%%%%%%%%%%%%%%%%%%%%%%%
%%%%%%%%%%%%%%%%%%%%%%%%%%%%%%%%%%%

%%%%%%%%%%%%%%%%%%%%%%%%%%%%%%%%%%%
%%%%%%%%%%%%%%%%%%%%%%%%%%%%%%%%%%%
\section{Evaluating the critical slope}
%for a defect pair, $\Pi_c^{(2)}$}
%{The critical capillary pressure slope for a single regular defect and a pair of identical defects}
\label{appendix_critical_slope}
%%%%%%%%%%%%%%%%%%%%%%%%%%%%%%%%%%%
%%%%%%%%%%%%%%%%%%%%%%%%%%%%%%%%%%%

Here, we derive the critical pressure slope for a \emph{pair} of identical defects, from which we obtain the result for a single defect by putting the two defects next to each other (equivalent to a single defect of twice the width), or infinitely far apart (equivalent to a single defect of the same width).
To analyze defects of an arbitrary width, we use the original differential equation for the interface shape, Eq.~\eqref{eq:press_balance}, instead of the Green's function-based approach.

Consider a pair of defects of width $w$, with distance $d$ between their symmetry axes, as introduced in Section~\ref{sec:cooperative_dissipation}. 
The capillary pressure profile is given by Eqs.~\eqref{eq:F2}, \eqref{eq:a2} and \eqref{eq:cy}. For the interface profile with $h_0=y_d$, if the interface solution {crosses the defects} entirely within the ramp regions, the equation for $\eta(x)=h(x)-y_d$ is
\begin{equation}
\gamma\eta''=\begin{cases}
(\rho g_e-\Pi)\eta, & ||x|-d/2|<w/2\ \text{(inside the defects)},\\
\rho g_e\eta, & ||x|-d/2|>w/2\ \text{(outside the defects)}.
\end{cases}
\end{equation}
For $\Pi>\rho g_e$ the solutions are given by
\begin{equation}
\eta(x)=\begin{cases}
C_1\cosh(x/\ell_c), & |x|<(d-w)/2,\\
C_2\cos[\sqrt{\Pi/\gamma-1/\ell_c^2}x]+C_3\sin[\sqrt{\Pi/\gamma-1/\ell_c^2}|x|], & (d-w)/2<|x|<(d+w)/2,\\
C_4\exp(-|x|/\ell_c), & |x|>(d+w)/2.
\end{cases}
\label{EqA2}
\end{equation}
The top line in \eqref{EqA2} corresponds to the space between the defects, the middle to the part of the interface inside the defects, and the bottom to the outer regions. 
Matching conditions, requiring that the values of $\eta(x)$ and its first derivative $\eta'(x)$ match at $|x|=(d-w)/2$ and $|x|=(d+w)/2$, provide a system of four linear homogeneous equations for four unknowns $C_1$--$C_4$, which has nontrivial solutions when its determinant is zero. 
Given the parameters $w$, $d$, $\ell_c$ and $\gamma$, this can only be satisfied for a single value of $\Pi=\Pi_c^{(2)}$. 
For this value, a continuum of nontrivial solutions exist, differing by the value of the prefactor from zero to the maximum value for which the solution {still crosses the defects entirely} within the ramps.

The interpretation of this result is as follows. 
(i) For $\Pi<\Pi_c^{(2)}$, only the trivial solution $\eta(x)=0$ of the original equation~\eqref{eq:press_balance} with $h_0=y_d$ exists; as $h_0$ increases, this solution evolves continuously, thus, the defect is weak. 
(ii) As $\Pi$ increases, the rate of deformation, $d\eta_m/dh_0$ 
    %(where $\eta_{\rm max}$ is the maximum value of $\eta$) \hl{I don't remember what we decided about this. We also introduced $\eta_m$, but is it the value on the defect axis or the maximum value? In the experiment, they are essentially the same, but in general this is not true. I've kept $\eta_{\rm max}$ here to avoid any ambiguity} 
grows, approaching infinity as $\Pi\to\Pi_c^{(2)}$ from below, so that at $\Pi=\Pi_c^{(2)}$ an infinitesimal change in $h_0$ gives rise to a finite deformation. 
This is consistent with the existence of a continuum of solutions with $h_0=y_d$ at $\Pi=\Pi_c^{(2)}$. 
(iii) for $\Pi>\Pi_c^{(2)}$ there are no nontrivial solutions that {cross the defects} within the ramps, but there is one {at least partially} within the {plateaus} and 
%\hl{note for a strong defect the interface can be partially in the ramp and partially in the plateau}, 
thus a jump occurs at $h_0=y_d$ and the defect is strong. 
Then $\Pi_c^{(2)}$ is the critical value separating weak and strong defect pairs.

By equating the determinant of the above-mentioned system of four equations to zero, an equation for $\Pi_c^{(2)}$ can be obtained. 
We define a quantity $\zeta$ such that
\begin{equation}
\label{eq:pic2vszeta}
\Pi_c^{(2)}=\gamma[1/\ell_c^2+(\zeta/w)^2].
\end{equation}
Then, the following equation
\begin{equation}
\label{eq:tanzeta}
\tan\zeta=\frac{1+\tanh[(d-w)/(2\ell_c)]}{(\ell_c\zeta/w)^2-\tanh[(d-w)/(2\ell_c)]}\frac{\ell_c\zeta}{w}
\end{equation} 
has one solution in the interval $0<\zeta<\pi$, which, generally, needs to be found numerically, and then $\Pi_c^{(2)}$ is given by Eq.~\eqref{eq:pic2vszeta}. 
If $\Pi$, $w$, $\ell_c$ and $\gamma$ are given, a fully analytical solution for the critical value of the distance $d$ is possible.

Next, we verify that for narrow defects ($w/\ell_c\ll 1$), the result of Eq.~\eqref{ac} is recovered. 
Assuming (to be confirmed by the calculation) that $w/\ell_c\ll\zeta\ll 1$, we can approximate Eq.~\eqref{eq:tanzeta} as
\begin{equation}
\zeta=\{1+\tanh[d/(2\ell_c)]\}\frac{w}{\ell_c\zeta},
\end{equation}
the solution of which is
\begin{equation}
\zeta=\left[\frac{w}{\ell_c}\left(1+\tanh\frac{d}{2\ell_c}\right)\right]^{1/2}.
\end{equation}
This indeed satisfies the above inequality for $\zeta$. Then, we write Eq.~\eqref{eq:pic2vszeta},
\begin{equation}
\Pi_c^{(2)}=\gamma\left[\frac{1}{\ell_c^2}+\frac{1+\tanh(d/2\ell_c)}{w\ell_c}\right]\approx\frac{\gamma[1+\tanh(d/2\ell_c)]}{w\ell_c},
\end{equation}
which, after a simple transformation, coincides with Eq.~\eqref{ac}. 
On the other hand, for $w\gg\ell_c$, since $\zeta$ is finite, Eq.~\eqref{eq:pic2vszeta} gives $\Pi_c^{(2)}=\gamma/\ell_c^2$ for any $d$.

Finally, we obtain the critical slope for a single defect by noting that for $d=w$ (two defects put together with no gap), we get
\begin{equation}
\tan\zeta=\frac{w}{\ell_c\zeta}.
\end{equation}
This corresponds to a single defect of width $2w$; then, for a single defect of width $w$,
\begin{equation}
\label{eq:tanu}
\tan u=\frac{w}{2\ell_c u},
\end{equation}
and
\begin{equation}
\label{eq:pic1vsu}
\Pi_c^{(1)}=\gamma[1/\ell_c^2+(2u/w)^2].
\end{equation}
Equation~\eqref{eq:tanu} is a transcendental equation for $u$ that needs to be solved numerically. 
However, a fully analytic solution is possible for the critical value of $w$ given $\Pi$, $\ell_c$ and $\gamma$.
%if the goal is to find the critical value of $w$ given $\Pi$, $\ell_c$ and $\gamma$, then a fully analytic solution is possible.
%
We can also check that for two defects very far apart ($d-w\gg\ell_c$) this single-defect result is recovered. 
Indeed, in this case Eq.~\eqref{eq:tanzeta} becomes
\begin{equation}
\tan\zeta=\frac{2\ell_c\zeta/w}{(\ell_c\zeta/w)^2-1}=\frac{2(w/\ell_c\zeta)}{1-(w/\ell_c\zeta)^2}.
\end{equation}
Using the trigonometric identity $\tan 2X=2 \tan{X}/(1-\tan^2X)$, we get
\begin{equation}
\tan(\zeta/2)=\frac{w}{\ell_c\zeta},
\end{equation}
which coincides with Eq.~\eqref{eq:tanu} if $u=\zeta/2$; Eq.~\eqref{eq:pic1vsu} then coincides with Eq.~\eqref{eq:pic2vszeta}.

%%%%%%%%%%%%%%%%%%%%%%%%%%%%%%%%%%%
%%%%%%%%%%%%%%%%%%%%%%%%%%%%%%%%%%%
\section{Mixed numerical-analytical computation of the interface profile}
\label{appendix_profiles}
%%%%%%%%%%%%%%%%%%%%%%%%%%%%%%%%%%%
%%%%%%%%%%%%%%%%%%%%%%%%%%%%%%%%%%%

Here, we describe the mixed numerical-analytical method we have used to calculate the interface configuration $h(x)$, in particular, (i) after the jump during imbibition and (ii) before the jump during drainage. 
As similar approaches have been used for a single and a pair of defects, we describe both cases at the same time, indicating differences where applicable.

We consider capillary pressure profiles given by Eqs.~\eqref{pcxz}--\eqref{eq:cy} for a single defect, and Eqs.~\eqref{eq:F2}, \eqref{eq:a2}, \eqref{eq:cy} for a pair. 
Thus, the defect consists of a ramp (slope) of length $\ell_s$, followed by a plateau. 
We only consider cases where in the range(s) of $x$ where the defect(s) is (are) located, the interface is entirely within the defect(s), i.e. for $|x|<w/2$ (single defect) or $||x|-d/2|<w/2$ (pair), $y_d<h(x)<y_d+\ell$. 
The interface then obeys the following equation for a single defect,
\begin{equation}
\label{eq:hsingle}
\gamma h''=\begin{cases}
\rho g_e(h-h_0), & |x|>w/2,\\
\rho g_e(h-h_0)-\Pi (h{-y_d}), & |x|<w/2\ \text{and}\ h<y_d+\ell_s,\\
\rho g_e(h-h_0)-\Pi (\ell_s{-y_d}), & |x|<w/2\ \text{and}\ h>y_d+\ell_s
\end{cases}
\end{equation}
and for a defect pair,
\begin{equation}
\label{eq:hpair}
\gamma h''=\begin{cases}
\rho g_e(h-h_0), & |x|>(d+w)/2,\ |x|<(d-w)/2\\
\rho g_e(h-h_0)-\Pi (h{-y_d}), & ||x|-d/2|<w/2\ \text{and}\ h<y_d+\ell_s,\\
\rho g_e(h-h_0)-\Pi (\ell_s{-y_d}), & ||x|-d/2|<w/2\ \text{and}\ h>y_d+\ell_s.
\end{cases}
\end{equation}

In the outer region ($|x|>w/2$ for a single defect and $|x|>(d+w)/2$ for a pair), the solution is
\begin{equation}
h=h_0+\eta_b\exp(-\Delta x/\ell_c),
\end{equation}
where $\Delta x=|x|-w/2$ for a single defect or $\Delta x=|x|-(d+w)/2$ for a pair. 
The constant $\eta_b$, the value of $h$ at the (outer) boundary of the defect, needs to be found based on the requirement that the solution is symmetric, thus, $h'(0)=0$, by matching to other parts of the solution, as discussed below; the solution for $\eta_b$ may or may not exist depending on $h_0$ and the parameters of the defect(s).

Inside the defect(s) ($|x|<w/2$ for a single defect or $||x|-d/2|<w/2$ for a pair), the solution, in general, consists of pieces of functions that can be found analytically.
As the matching between these pieces is cumbersome, that part of the solution is obtained here by numerical integration. 
Assuming that $h_0$ and $\eta_b$ are known, the values of $h$ and its first derivative are provided from the outer solution at the $x=w/2$ [or $x=(d+w)/2$] boundary of the defect, and can serve as the initial conditions for numerical integration. 
For a single defect, integration {can be} carried out down to $x=0$ to find out if the $h'(0)=0$ condition is satisfied. 
{Thus, we can} find $\eta_b$ by solving the $h'(0)=0$ equation using the bisection root-finding scheme~\cite[Chapter 9.1]{NumericalRecipes}. 
For a pair of defects, $x=0$ is in the middle between the defects; in that region between the defects, the analytical solution with $h'(0)=0$ is
\begin{equation}
h(x)=h_0+C\cosh(x/\ell_c),
\end{equation}
where $C$ is an unknown constant. 
This gives the condition on the values of $h$ and its first derivative on the inner boundary of the defect,
\begin{equation}
\label{eq:ratiocond}
\frac{h'([d-w]/2)}{h([d-w]/2)-h_0}=\frac{1}{\ell_c}\tanh\frac{d-w}{2\ell_c}.
\end{equation}
By integrating numerically down to $(d-w)/2$ within the defect, we find if this condition is satisfied, which, as before provides an equation for the bisection scheme to find $\eta_b$.
We note that for $\Pi>\rho g_e$ the solution {for $|x|<w/2$ (or $||x|-d/2|<w/2$)} can be oscillatory and it is possible that several roots $\eta_b$ and associated solutions exist. 
However, apart from {the solution with a single maximum inside (each) defect}, these solutions go below $y_d$ and as such violate the conditions outlined above.

During imbibition, the interface jump happens once the interface touches the defect. 
Thus, $h_{\rm imb}^{-}=y_d$ and $h_{\rm imb}^{+}$ is the solution of Eq.~\eqref{eq:hsingle} or \eqref{eq:hpair} with $h_0=y_d$. 
This solution can be found as described above. 
For drainage, the situation is more complicated, because $h_0$ is unknown. 
Its value is a bifurcation point such that for lower values there are no solutions {crossing} the defect. 
In other words, for lower $h_0$, regardless of the value of $\eta_b$, the defect {will} not sufficiently deform the interface. 
For a single defect this means that the maximum possible value of $h'(0)$ is below zero; for a pair, the maximum possible value of the ratio on the left-hand side of Eq.~\eqref{eq:ratiocond} is smaller than its right-hand side. 
We then carry out a nested procedure, where in the inner cycle, for a particular $h_0$ we find the maximum value of $h'(0)$ (for a single defect) or of the left-hand side of Eq.~\eqref{eq:ratiocond} (for a pair), using the golden-section algorithm~\cite[Chapter {10.2}]{NumericalRecipes}, and then, via bisection, find the value of $h_0$ for which this maximum value is zero (for a single defect) or the right-hand side of Eq.~\eqref{eq:ratiocond} (for a pair). 
This eventually provides both $h_{\rm dra}^{-}$ and $h_{\rm dra}^{+}$ (the latter equal to the value of $h_0$ resulting from the procedure).

\begin{figure}
\includegraphics[width=.45\textwidth]{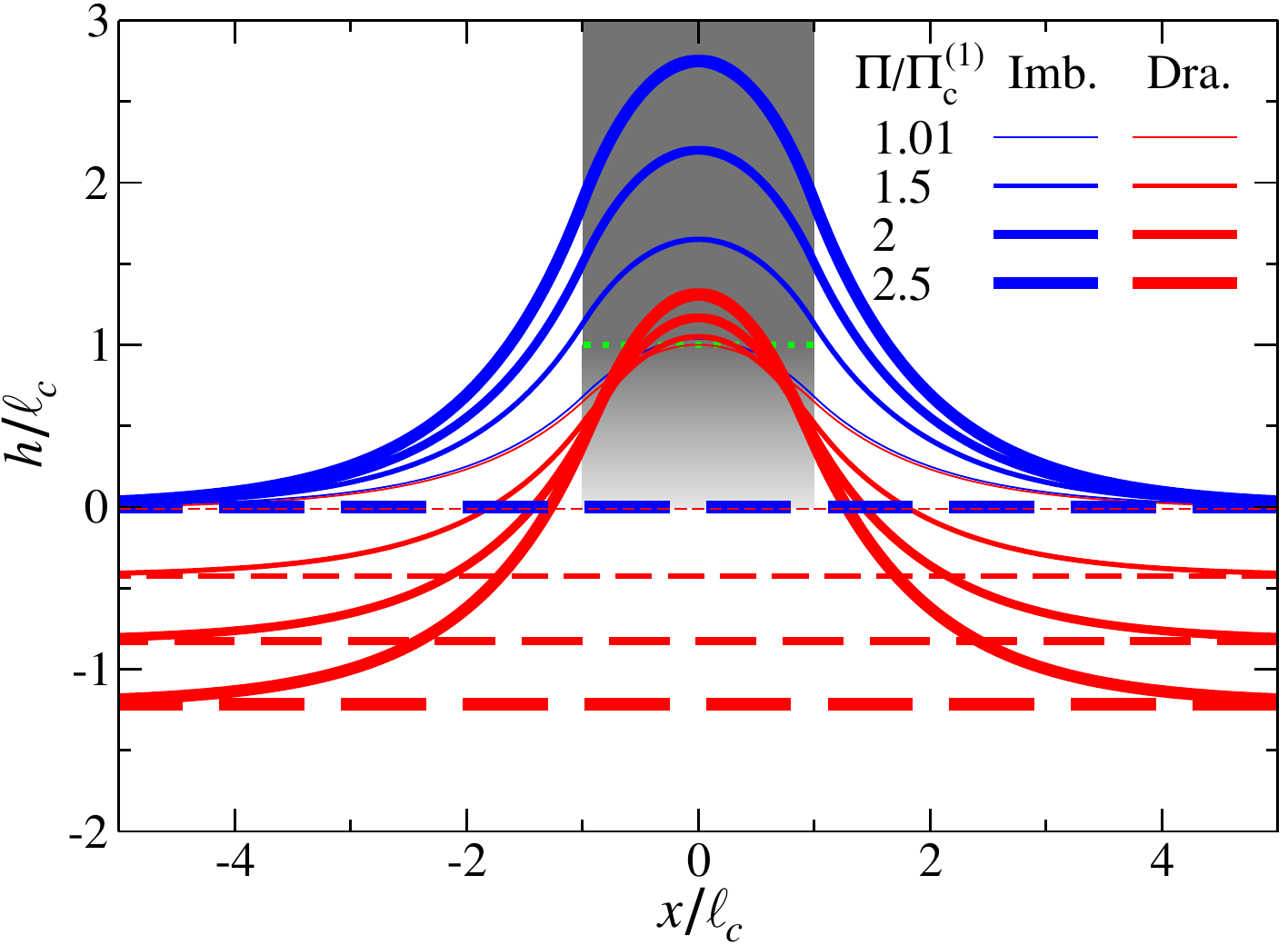}
            \centering
\caption{
Interface profiles before and after jumps for an intermediate-width ($w/\ell_c = 2$) regular defect, for different capillary pressure slopes. The ramp part of the defect is shown as gradient shading and the plateau part in uniform gray. The boundary between these two regions at $h/\ell_c=1$ is marked with a green dotted line. Dashed and solid lines represent the undeformed (flat) and deformed configurations, i.e. before and after the jump in imbibition, and the opposite in drainage.
These results are complementary to those shown in Fig.~\ref{Fig:profiles}.
}
\label{Fig:profiles_interm}
\end{figure}

The above procedure was used to find the interface shapes for narrow and wide single defects in Fig.~\ref{Fig:profiles}. 
For clarity, we add a similar plot for the general case of a defect of an intermediate width ($w/\ell_c=2$; Fig.~\ref{Fig:profiles_interm}). 
{Overall}, qualitatively, the results are intermediate between the limits of wide and narrow defects. 
However, notably, for imbibition, while in both limits $\eta(0)/\ell_s=\Pi/\Pi_c^{(1)}$, here the value of $\eta(0)$ is slightly higher, thus, the dependence of $\eta(0)$ on the defect width is non-monotonic.

Once the interface profiles before and after the jump are found, the dissipated energy can be calculated using Eqs.~\eqref{eq:diss_dif} and \eqref{eq:diss_dra}, where the integration is carried out numerically inside the defect(s) and analytically outside. 
{Note that if one fixes $\eta_b$ (for imbibition) or $h_0$ (for drainage), the corresponding value of $d$ for which Eq.~\eqref{eq:ratiocond} is satisfied can be found directly, avoiding bisection. This can be used to speed up computations when obtaining
$\Psi$ vs $d$ data (Fig.~\ref{Fig:weak_pair}), by varying $\eta_b$ (or $h_0$) and producing a table of [$\Psi$, $d$] values.}

%%%%%%%%%%%%%%%%%%%%%%%%%%%
%%%%%%%%%%%%%%%%%%%%%%%%%%%
%%%%%%%%%%%%%%%%%%%%%%%%%%%
% \section{Derivations of analytical expressions}
% \label{appendix_analytical}

% \hl{RH: analytical expressions for what? can I ask whoever put this section to please elaborate, and fill or assign the right person to fill the text? thx}

% \section{Numerical simulations}
% \label{appendix_numerical}

% \hl{the below is copy paste from GRL SI on numerics: 1. change or refer to GRL? 2. add a bit on error, num dissipation.}

%%%%%%%%%%%%%%%%%%%%%%%%%%%
%%%%%%%%%%%%%%%%%%%%%%%%%%%
%%%%%%%%%%%%%%%%%%%%%%%%%%%

%%%%%%%%%%%%%%%%%%%%%%%%%%%
%%%%%%%%%%%%%%%%%%%%%%%%%%%
%%%%%%%%%%%%%%%%%%%%%%%%%%%
\section{Numerical details}
\label{appendix_numerical}
%%%%%%%%%%%%%%%%%%%%%%%%%%%
%%%%%%%%%%%%%%%%%%%%%%%%%%%
%%%%%%%%%%%%%%%%%%%%%%%%%%%
% RH 150123 moved to APPENDIX
For the results shown in Fig.~\ref{Fig:weak_pair} (and Videos S2--S4 in {\color{blue}
%\href{https://www.dropbox.com/scl/fo/rkfy7j5xr8ofqr76m7avi/h?rlkey=cmgixv6dhfh01s9b2p85qrn5e&dl=0}
{SI}}), we use the following parameters: $g_e = 0.2$ m/s$^2$, $\rho = 998$ kg/m$^3$, $\gamma = 20.7$~mN/m (such that $\ell_c \approx 10.2$~mm). 
The defect width is $w=0.8$~mm, which means that it is relatively narrow ($w / \ell_c \approx 0.08$), and we expect the narrow defect theory to yield a good approximation. 
The defect profile is such that $\delta p_c$ changes linearly between zero and $\approx$$7.39$ Pa along a slope of length $\ell_s = 2$~mm. 
For these values, an isolated single defect is weak, as the pressure slope, $\Pi\approx 3.70\times 10^3$~Pa/m, is smaller than the critical value for a single defect, computed {using} both (i) the narrow-defect approximation, with Eq.~\eqref{eq:crit_slope_weak_single} giving $\Pi_n^{(1)} \approx 5.08\times 10^3$~Pa/m; and (ii) {the exact arbitrary-width result of Appendix~\ref{appendix_critical_slope}}, with Eqs.~\eqref{eq:tanu}--\eqref{eq:pic1vsu}, such that $\Pi_c^{(1)}\approx 5.22\times 10^3$~Pa/m.

%%%%%%%%%%%%%%%%%%%%%%%%%%%
%%%%%%%%%%%%%%%%%%%%%%%%%%%
%%%%%%%%%%%%%%%%%%%%%%%%%%%
\section{Experimental details}
\label{appendix_experimental}
%%%%%%%%%%%%%%%%%%%%%%%%%%%
%%%%%%%%%%%%%%%%%%%%%%%%%%%
%%%%%%%%%%%%%%%%%%%%%%%%%%%

\begin{figure}
\includegraphics[width=0.75\columnwidth,keepaspectratio]{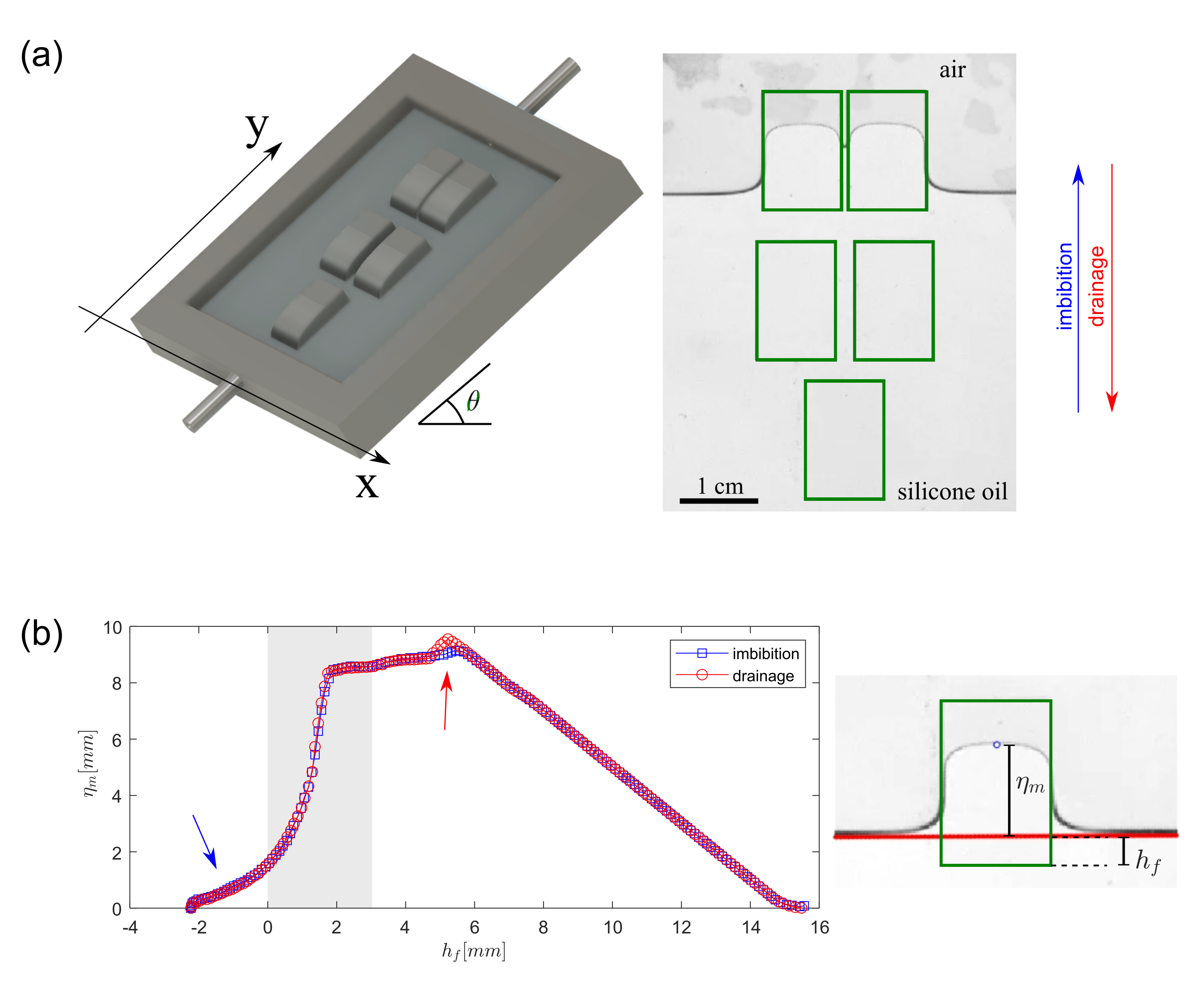}
\caption{(a) A 3-D printed 
%model of the imperfect Hele-Shaw cell 
including the 3 sets of defects (left), with a typical image of the interface during an experiment (right), showing the interface deformation due to its interaction with the top defects (dark green rectangles). 
Silicone oil enters from the bottom to displace the air initially filling the cell. 
The oil-air interface advances (imbibition) and recedes (drainage) in response to an increase or decrease in the oil pressure at the bottom of the cell, respectively. 
The cell is tilted by an angle $\alpha = 38^{\circ}$ with respect to the horizontal. (b) Imbibition-drainage cycle for the single defect. For each frame analyzed, we track the unperturbed position of the interface far from the defect $h_f$ and the maximum perturbation $\eta_m$ of the interface with regards to the unperturbed position. 
The presented cycle is reversible, namely the defect is \emph{weak}. 
In panel (b), the gray shaded area corresponds to the data shown in Fig.~\ref{fig:hysteresis} (a), where the arrows point to interesting physical effects not considered in our theoretical analysis (see text).}
\label{fig:3D_model}
\end{figure}

An imperfect Hele-Shaw cell was manufactured using stereolithography 3-D printing.
%Figure~\ref{fig:3D_model} shows a sketch of the setup. 
The cell was produced with three sets of defects: a single defect and two sets of defect pairs with varying inter-defect gaps, see Fig.~\ref{fig:3D_model} (a). This efficient design allows us to run three separate experiments on the same cell in sequence. 

To experimentally {produce} variations in the capillary pressure $\delta p_c$, we introduce modulations of the Hele-Shaw cell gap space $b(x,y) = b_0-\delta b(x,y)$. This introduces a local variation of the out-of-plane capillary pressure $p_c(x,y) = p_c^{0} + \delta p_c(x,y)$, where $p_c^0 = 2 \gamma \cos(\theta)/b_0$ ($\theta$ being the contact angle)
and $\delta p_c = p_c^0 \delta b/(b_0 - \delta b)$~\cite{HoltzmanCommPhys2020}.
%
%As a first order approximation, we can consider the capillary pressure along the defect to be given by $p_c(x,y) = 2\gamma/(b_0-\delta b)$. 
We design the functional form of $\delta b(y)$ such that the the capillary pressure $p_c(x,y)$ {within the defect} is a linear function of $y$, i.e., $\delta p_c(x,y) = \Pi (y-y_d)$ for 
% $0<y\leq l$,
$y_d<y\leq y_d+\ell_s$,
and $\delta p_c(x,y) = \Pi\ell_s$ for $y_d+\ell_s < y < y_d+\ell$. 
The length of the sloping part of the defect is $\ell_s=10$~mm 
and $\ell_p=\ell-\ell_s=5$~mm is the length of the constant {capillary} pressure zone {(plateau)} after the slope. 
The defect width is $w = 10$~mm. The value for the pressure slope $\Pi = dp_c/dy$ was chosen through a series of experiments as $\Pi = 6\times 10^3$~Pa/m. 
This value was chosen as it provides interface deformations that are large enough to be easily captured by image analysis, while not too large to avoid highly nonlinear deformations and snap-off events during drainage.
%was found suitable to for the particular feature we want to investigate in this study (the cooperative dissipation arising from the interaction of defects). 
Under these conditions, the defect profile is:
$\delta b(y) = b_0 - \left(\frac{\Pi (y -y_d)}{2\gamma}+\frac{1}{b_0}\right)^{-1}$
%
% \begin{equation}
%     \delta b(y) = b_0 - \frac{1}{\frac{Ay}{2\gamma}+\frac{1}{b_0}}
% \end{equation}
for $y_d<y\leq y_d+\ell_s$, 
and 
%
% \begin{equation}
% \delta b(y) = \delta b(l)
% \end{equation}
$\delta b(y) = \delta b(\ell_s)$
for $y_d+\ell_s < y < y_d+\ell_s+\ell_p$ (here we made the assumption that the liquid perfectly wets the medium).
The cell's area is 6 cm $\times$ 6 cm, with a depth of $b_0 = 3.6$~mm. For wetting fluid we used {silicone} oil, with kinematic viscosity $\nu = 10$ cSt $= 10$ mm$^2$/s, surface tension against air $\gamma = 20$~mN/m and density $\rho = 0.93$~g/mL. The non-wetting fluid is ambient air at atmospheric pressure. 

The experiment is driven by changing the height of a reservoir of silicone oil connected to the inlet (bottom) of the model. As the reservoir height increases by $\delta h$, the oil pressure at the bottom of the model increases by $\delta p = \rho g \sin(\alpha) \delta h$ thus causing the interface to move upwards (imbibe), and vice versa in drainage.
A moving average filter is applied to the $h_f$ and $\eta_m$ data (in Figs.~\ref{fig:hysteresis} and~\ref{fig:3D_model}) to remove spurious high-frequency noise (a consequence of image analysis inaccuracies).

The experiments in Fig.~\ref{fig:hysteresis} \emph{qualitatively} demonstrate that (1) the imbibition-drainage cycle around a single defect can be reversible, thus characterizing a weak defect, and (2) the interaction between weak defects can trigger irreversibility (hysteresis). 
The experiments also reveal further intriguing physics not considered in our model. 
This is evident in the reversible case shown in Fig.~\ref{fig:3D_model}b: 
Examining the entire curve in Fig.~\ref{fig:3D_model}b, outside the gray shaded area which corresponds to the region in Fig.~\ref{fig:hysteresis} where potential hysteresis and dissipation may occur according to our model (see Fig.~\ref{Fig2:pressure_balance}), exposes other interesting features. 
The blue arrow in Fig.~\ref{fig:3D_model}b points to a perturbation in the curve even before the baseline position reaches the defect. 
This can be explained through the 3-D nature of the liquid-air interface. 
As the silicone oil wets both the top and bottom surfaces of the cell, there is a thin film of liquid ahead of the 2-D projected interface. 
Once this film touches the defect, the perturbation starts to grow even before its 2-D projection touches the defect. 
Similarly, the red arrow in Fig.~\ref{fig:3D_model}b points to a small bump in the drainage cycle, which might be caused by pinning of the contact line as it touches the defect in drainage. 
These 3-D effects are not included in our 2-D model. The plateau region included after the sloping part of the defect ensures that artifacts such as those related to contact line pinning (red arrow in Fig.~\ref{fig:3D_model}b) occur far from the area of interest in our model (gray shading in Fig.~\ref{fig:3D_model}b).

%%%%%%%%%%%%%%%%%%%%%%%%%%%%%%%%
\begin{acknowledgments}
%\noindent{\textbf{Acknowledgments}: 
RH acknowledges support from the Engineering and Physical Sciences Research Council (EP/V050613/1); MD and JO received support from the Spanish Ministry of Science and Innovation through the project HydroPore (PID2019-106887GB-C31 and C32, PID2022-137652NB-C41 and C42), and RP through the project PID2021-122369NB-I00. RP and JO acknowledge AGAUR (Generalitat de Catalunya) for financial support through project 2021-SGR-00450. MM acknowledges the support from the Research Council of Norway through projects 262644 and 324555.
\end{acknowledgments}
%%%%%%%%%%%%%%%%%%%%%%%%%%%%%%%%

%%% ARXIV %%% \bibliography{bib_dissipation_detailed_paper.bib}

%%% ARXIV 
%apsrev4-2.bst 2019-01-14 (MD) hand-edited version of apsrev4-1.bst
%Control: key (0)
%Control: author (8) initials jnrlst
%Control: editor formatted (1) identically to author
%Control: production of article title (0) allowed
%Control: page (0) single
%Control: year (1) truncated
%Control: production of eprint (0) enabled
%

\end{document}